\documentclass[a4paper,12pt]{article}
\pdfoutput=1
\usepackage{cite}
\usepackage{amsmath}
\usepackage{amsfonts}
\usepackage{amssymb}
\usepackage{graphicx, rotating}
\usepackage{epsfig}
\usepackage{latexsym}
\usepackage{graphicx}
\usepackage{color}
\usepackage{amsmath,bm,amssymb}

\usepackage{color}
\usepackage[english]{babel}
\usepackage{hyperref}

\newcommand{\Slash}[1]{{\ooalign{\hfil#1\hfil\crcr\raise.167ex\hbox{/}}}}

\newcommand{\beq}{\begin{equation}}  \newcommand{\eeq}{\end{equation}}
\newcommand{\bef}{\begin{figure}}  \newcommand{\eef}{\end{figure}}
\newcommand{\bec}{\begin{center}}  \newcommand{\eec}{\end{center}}
  
\newcommand{\laq}[1]{\label{eq:#1}}  

\newcommand{\Eq}[1]{Eq.~(\ref{eq:#1})}
\newcommand{\Eqs}[1]{Eqs.~(\ref{eq:#1})}
\newcommand{\eq}[1]{(\ref{eq:#1})}
\newcommand{\Sec}[1]{Sec.\ref{chap:#1}}
\newcommand{\ab}[1]{\left|{#1}\right|}
\newcommand{\vev}[1]{ \left\langle {#1} \right\rangle }

\newcommand{\lac}[1]{\label{chap:#1}}
\newcommand{\SU}[1]{{\rm SU{#1} } }

\def\({\left(}
\def\){\right)}

\def\O{\mathcal{O}}
\def\U{\mathop{\rm U}}

\def\ebq{\end{equation} \begin{equation}}
\newcommand{\OR}{~{\rm or}~}
\newcommand{\AND}{~{\rm and}~}

\newcommand{\KEV}{ {\rm \, keV} }
\newcommand{\MEV}{ {\rm \, MeV} }
\newcommand{\GEV}{ {\rm \, GeV} }

\def\a{\alpha}

\def\d{\delta}
\def\e{\epsilon}
\def\f{\phi}

\def\h{\theta}

\def\r{\rho}

\def\L{\Lambda}
\def\F{\Phi}

\def\tl{\tilde}
\def\*{\dagger}

\begin{document}
\pagestyle{empty}
\renewcommand\bibname{\Large References}

\begin{center}

\hfill   KEK-TH-2310\\

\vspace{1.5cm}

{\Large\bf Strong CP problem and axion dark matter \\ with small instantons}\\

\vspace{1.5cm}

{Ryuichiro Kitano$^{1,2}$ and Wen Yin$^{3}$}

\vspace{12pt}
{\em 
$^1$KEK Theory Center, Tsukuba 305-0801, Japan\\
$^2$Graduate University for Advanced Studies (Sokendai),
Tsukuba 305-0801, Japan\\
$^3${Department of Physics, Faculty of Science, The University of Tokyo,\\
Bunkyo-ku, Tokyo 113-0033, Japan} \vspace{5pt}} \\

\vspace{1.5cm}
\abstract{The axion mass receives a large correction from small
instantons if the QCD gets strongly coupled at high energies. We
discuss the size of the new CP violating phases caused by the fact
that the small instantons are sensitive to the UV physics. We also
discuss the effects of the mass correction on the axion abundance of
the Universe. Taking the small-instanton contributions into account,
we propose a natural scenario of axion dark matter where the axion
decay constant is as large as $10^{15\text{-}16}$\,GeV. The
scenario works in the high-scale inflation models.
 
 }

\end{center}
\clearpage

\setcounter{page}{2}
\setcounter{footnote}{0}
\pagestyle{plain}

\section{Introduction}

The axion, a hypothetical light particle couples to QCD and QED,
drastically modifies physics at long distance scales. Especially, when
its mass is dominated by the contributions from the QCD, the vacuum is
chosen to eliminate the strong CP phase, thereby solving the strong CP
problem in the standard model of particle
physics~\cite{Peccei:1977hh,Peccei:1977ur,Weinberg:1977ma,Wilczek:1977pj}.
The existence of such a new degree of freedom is also motivated as a
candidate of dark matter of the Universe~\cite{Abbott:1982af,Preskill:1982cy, Dine:1982ah}.
The axion can be originated from various microscopic models, for
example, as the Nambu-Goldstone particle associated with the
Peccei-Quinn (PQ)
symmetry~\cite{Peccei:1977hh,Peccei:1977ur,Weinberg:1977ma,Wilczek:1977pj}
and also as a part of theories of quantum gravity~\cite{Witten:1984dg,Svrcek:2006yi,Conlon:2006tq,Arvanitaki:2009fg,Acharya:2010zx,Higaki:2011me, Cicoli:2012sz,Demirtas:2018akl} (see
Refs.\,\cite{Jaeckel:2010ni,Ringwald:2012hr,Arias:2012az,Graham:2015ouw,Marsh:2015xka,Irastorza:2018dyq, DiLuzio:2020wdo} for reviews.)

An essential feature of the axion in order to solve the strong CP
problem is the shift symmetry, $a \to a + c$, which is only broken
by the non-perturbative dynamics of QCD through the coupling to the
instanton density in the Lagrangian.
The vacuum is automatically CP conserving once this system is realized
as the low energy effective theory.
The shift symmetry is naturally realized as the Nambu-Goldstone
mechanism of the spontaneously broken PQ symmetry.
Although it sounds like a good solution to the strong CP problem, the
requirement of the shift symmetry poses another question that why or
how the PQ symmetry, which is anomalous to QCD, is maintained with a
great accuracy in fundamental theories to UV complete the effective
theory.
Indeed, violation of the PQ symmetry is generally present in theories
of quantum gravity~\cite{Misner:1957mt, Barr:1992qq,Banks:2010zn,Alvey:2020nyh}.
Field theoretic model building to resolve this ``quality problem'' has
been discussed in literature. One approach is to regard the PQ
symmetry as an accidental symmetry protected by some gauge
symmetries~\cite{Chun:1992bn,BasteroGil:1997vn,Babu:2002ic}\cite{Fukuda:2017ylt,Duerr:2017amf,Bonnefoy:2018ibr}\cite{Randall:1992ut,DiLuzio:2017tjx,Lillard:2018fdt,
Lee:2018yak, Ardu:2020qmo, Yin:2020dfn, Yamada:2021uze}. (See also
Ref.\,\cite{Redi:2016esr} and
Refs.\,\cite{Darme:2021cxx,Nakai:2021nyf}).
Another interesting approach is to identify the axion as a part of
gauge fields in larger space-time
dimensions~\cite{Cheng:2001ys,Choi:2003wr,Acharya:2010zx,Marsh:2019bjr}.

Once the effective theory with approximate shift symmetry is realized,
there is a relation between the axion mass $m_a$ and the decay
constant $f_a$ as $m_a f_a \sim (100~{\rm MeV})^2$. The axion
scenarios with this relation has a particular name, the QCD axions,
and have been distinguished from other axion-like particles which do
not solve the strong CP problem.
Recently, however, there have been extensive discussions on the
possibility of heavier axions than the QCD ones while the strong CP
problem is still solved.
Such a scenario is possible if there is some UV dynamics which makes
the QCD get strong again at high energy. In this case, the instanton
configurations with small sizes can give a large contribution to the
axion mass while it is aligned to the low energy contributions since
it is still the QCD effects.
Examples to realize such UV strong QCD is to embed the $\SU(3)$ gauge
group to a larger gauge group at a high energy
scale~\cite{Agrawal:2017ksf, Csaki:2019vte,Gherghetta:2020ofz,Gupta:2020vxb} and also to let gluons propagate
into a small extra dimension~\cite{Poppitz:2002ac,Gherghetta:2020keg}. 
The enlarged parameter space of the QCD axions will be quite important
for low energy axion phenomenology, axion cosmology, and also for
axion astrophysics.

In this paper, we consider how the small instantons affect the axion
potential in the current Universe, in the early Universe, and during
the inflation.
In particular, if the contributions from small instantons are
important, CP phases in dimension six operators in the standard model
effective theory cause a misalignment of the vacuum from the CP
preserving one, and thus reintroduce the strong CP problem.
Taking this effects into account, we discuss how large the UV
contributions can be in general setups. Note that the misalignment
caused by the small instantons is independent of the quality problem
of the PQ symmetry as the dimension six operators to cause the problem
are invariant under the PQ symmetry.

In the presence of the small instanton contributions to the axion
mass, the axion abundance generated by the misalignment mechanism is
modified. There are parameter regions where the axion abundance is
reduced, which allows a larger axion decay constant.

The small instantons are particularly important in the models where
the axion arises from a gauge field in the extra dimension such as
string axions.
The small instantons which stretch over the extra dimension is
considered in Ref.~\cite{Gherghetta:2020keg} and it is found that the axion mass can be much
heavier than that in the conventional scenario by many orders of
magnitude.
However, by considering the CP phases in the higher dimensional
operators, such an enhancement of the axion mass should be avoided.
We find that a consistent scenario requires the size of the extra
dimension, i.e., the axion decay constant to be larger than about
$10^{15-16}$~GeV where the UV contributions to the axion mass is much
smaller than the conventional low energy contribution.
Although the strong CP problem can be avoided, such a light axion is
cosmologically severely constrained by the isocurvature of the density
perturbation and the overproduction of the axion by the misalignment
mechanism.

The small instanton effects, however, provides us with an interesting
cosmological scenario in the extra-dimensional model.
One can consider the possibility that the radius of the extra
dimension during the inflation is smaller than the current size, i.e.,
the grand unification scale. Such a situation can be easily realized
when the volume modulus (radion) is sufficiently light or if the radion is the inflaton
itself.
In this case, the QCD scale gets higher during the inflation and the
axion field can be stabilized near the CP preserving point without
introducing the isocurvature perturbations.
A small displacement caused by the CP violating small instanton
effects can explain the correct abundance for the axion dark matter.

\section{Small instantons and CP problem}
In this section, we study the small instanton contribution to the axion potential by taking account of various higher dimensional terms with CP-phases. 
We will see that such contributions generically shift the minimum of the QCD axion potential to a CP-violating position and thus reintroduce the strong CP problem. 
We discuss the relation to the quality problem of the PQ symmetry.

\subsection{Aligned axion potential from small instantons}

\begin{figure}[!t]
\begin{center}  
   \includegraphics[width=105mm]{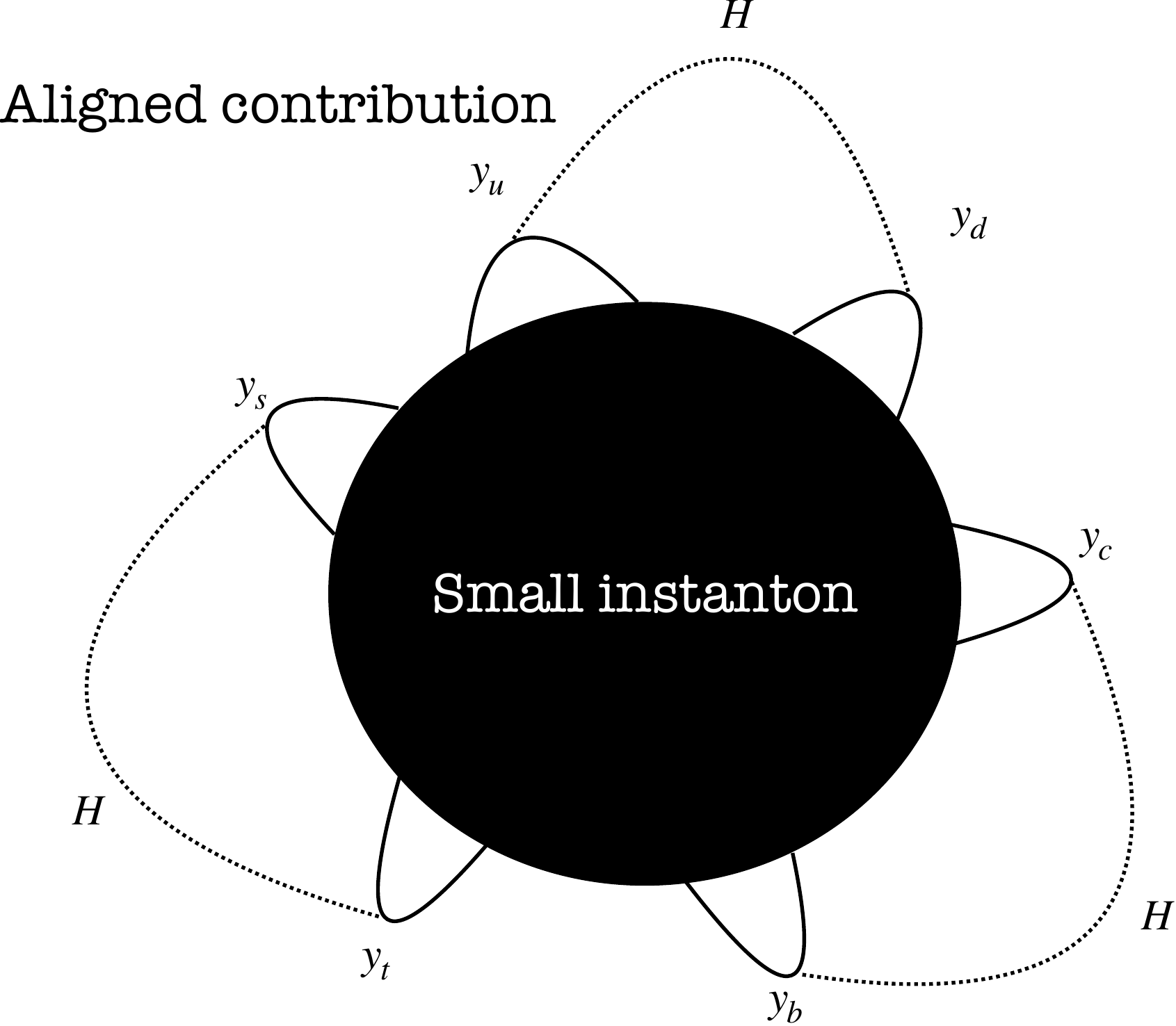}
      \end{center}
\caption{ Small instanton contribution to the QCD axion by assuming that only SM fermions are involved in the dynamics. The resulting potential has a CP phase aligned to the IR QCD potential.
}\label{fig:1} 
\end{figure}

We discuss contributions to the axion potential from small instantons
in general models.
We are particularly interested in the contributions with the instanton
sizes much smaller than the electroweak scale, and thus the vacuum
expectation value of the Higgs field can be ignored. The chirality flips
required to close the 't~Hooft vertex can be obtained by using the
Yukawa interactions, with the coupling constants $Y_u$ and $Y_d$, and
the loops of the Higgs lines as in Fig.~\ref{fig:1}.

By using the dilute instanton gas approximation~\cite{Callan:1977gz}, one can evaluate the
UV contribution to the axion potential~\cite{Holdom:1982ex, Dine:1986bg, Flynn:1987rs}

\beq
\laq{vac}
V (a)
= 
V_{\rm QCD} (a) + 
  \int^{1/\L_{\rm SM}}_{1/\L_{\rm cutoff}}{\frac{d \rho' }{\rho'^5} e^{-S_{\rm eff}[1/\rho']}  \
  F^{\rm (vac)}[\rho']
   \det{\frac{Y_{u}}{4\pi}} \det{\frac{Y_{d}}{4\pi}} } \cos{\(\frac{a}{f_a}\)}.
\eeq
Here, we separated the contribution from the infrared QCD dynamics,
$V_{\rm QCD}$.
The integration over $\rho'$ represents that of the size modulus of
the instanton solution. The dependence of the effective action on
$\rho'$ arises from the quantum corrections which are captured by the
running coupling constant, $S_{\rm eff}[1/\rho]\approx
2\pi/\a_s[1/\rho]$. The normalization of the instanton density is
evaluated as $F[\rho']\approx 10^{-3}\({2\pi / \a_s(1/\rho')}\)^6$ for
a single instanton~\cite{tHooft:1976snw}.
Depending on the behavior of the running gauge coupling, there can be
a large contributions from the second term.

The IR cut-off, $\Lambda_{\rm SM}$, can be arbitrary as the dependence
on $\Lambda_{\rm SM}$ is formally absorbed in $V_{\rm QCD}$.
The UV cut-off, $\Lambda_{\rm cutoff}$, represents the scale above
which we do not know the effective field theoretic description. If the
integral is dominated by the $\rho' \sim 1/\Lambda_{\rm cutoff}$ region,
the axion potential is not calculable within the effective theory
although the integral still provides an estimate.
We denote the typical instanton size which provides the largest
contributions to the integral as $\rho$ in the following discussions.

It is important to note that at this stage $V(a)$ and $V_{\rm QCD} (a)$
have the minimum at the same value of $a$ since there is no additional
CP phase in the discussion.
This aligned contributions can enhance the axion mass while the strong
CP problem is solved.
We denote this aligned contributions, i.e., the second term as
\beq
\laq{V1}
V_{\rm UV} (a)= \e \chi_0
\left( 1-\cos{
      \frac{a}{f_a}} \right),
\eeq
where $\chi_0$ is the topological susceptibility in QCD. The parameter
$\epsilon$ represents the relative size of the UV contribution to the
axion mass.

\subsection{CP violation from small instantons}

Any field theory involving gravity should be UV completed at a UV scale $\L$, and there should be
many higher dimensional terms~(c.f. \cite{Misner:1957mt, Barr:1992qq,Banks:2010zn,Alvey:2020nyh}).\footnote{If there were no higher
dimensional terms, the small strong CP phase is natural since it is
rarely generated via radiative correction. If there is, on the other
hand, $\O(1)$ phase is easily generated at the loop level, which makes
the strong CP problem really a problem.  } Thus, in general, we expect
CP violating terms originated from the UV physics, e.g. 
\beq
\laq{dim6}
{\cal L}\supset C_{ud}^{ijkl} \frac{  \bar{Q}_{i} u_j  \bar{Q}_k d_l }{\L^2}
\eeq
where $\L\gtrsim \L_{\rm cutoff}$ is the energy scale generating the operator, and 
$C_{ud}^{ijkl}$ a dimensionless coefficient.
The fields $Q$, $u$, and $d$ are quark fields in the standard model
with the generation indices.
Note that this term does not include the axion and do not violate the PQ symmetry.
We  expect 
\beq
C_{ud}^{ijkl}=\O(1)+ i \O(1),
\eeq
since gravity is argued to break any global symmetry. 
(We discuss the case where $C$ has chirality suppressions, i.e., $C_{ud}
\sim Y_u Y_d$ later.)
Here (and hereafter) for concreteness we have implicitly assumed that
 a KSVZ-like axion model~\cite{Kim:1979if,Shifman:1979if}, in which
 the quarks are not charged under the PQ symmetry, and thus \eq{dim6}
 is allowed by the PQ symmetry. In the case of DFSZ
 axion~\cite{Dine:1981rt,Zhitnitsky:1980tq}, on the other hand, this
 higher dimensional term is forbidden. Instead we can consider the
 terms such as $H_u Qu |H_u|^2/\L^2, (H_u H_d \F_{\rm PQ})^2/\L^2$
 with $H_u, H_d, \F_{\rm PQ}$ being the up-type, down-type and PQ
 Higgs fields, respectively. Our conclusions do not change
 qualitatively in these cases.

\begin{figure}[!t]
\begin{center}  
   \includegraphics[width=105mm]{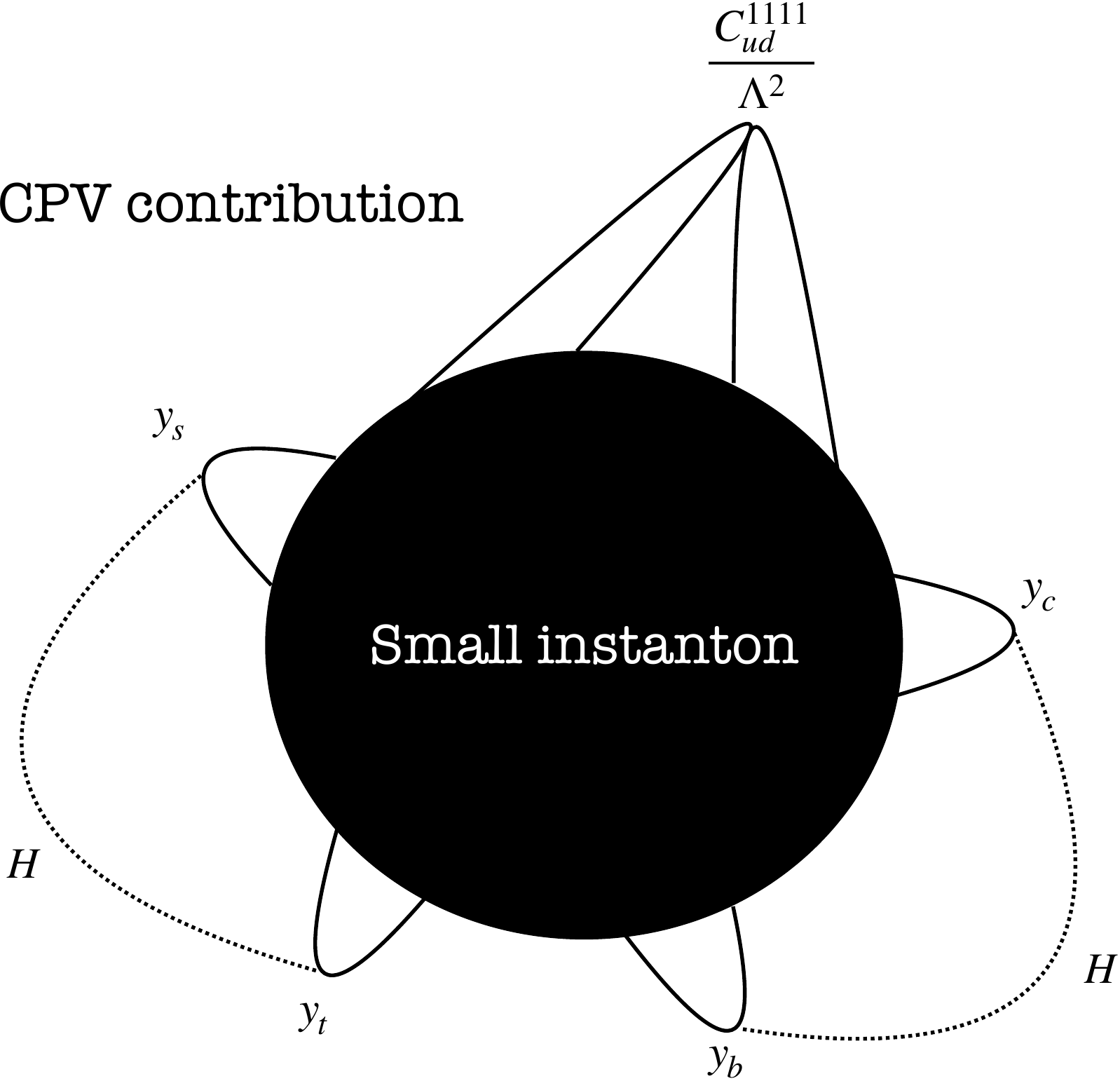}
      \end{center}
\caption{ One of the small instanton diagrams that gives a
new CP phase to the axion potential.}\label{fig:2} 
\end{figure}

Let us consider the small instanton contribution involving this term
via, e.g.,  the diagram in Fig.~\ref{fig:2}. 
The contribution is in general not aligned to $V_{\rm QCD} (a)$ such as 
\beq
\laq{CPV}
V_{\rm CPV} (a)
 \approx   \frac{\ab{C_{ud}^{1111}}}{(4\pi)^2\L^2} \frac{y_b y_s y_c y_t}{(4\pi)^4} \int_{1/\L_{\rm cutoff}}^{1/\L_{\rm SM}}{\frac{d \rho' }{\rho'^5}  \rho'^{-2} e^{-S_{\rm eff}[1/\rho']}  
F^{\rm (CPV)}(\rho')\cos{\(\frac{a}{f_a}+\arg{C}^{1111}_{ud}\)}}.
\eeq
where we have neglected the CKM matrix for simplicity of notation, and
we have used that $C_{ud}^{1111}$ is the dominant since the SM Yukawa
couplings satisfy $y_u, y_d\ll y_s, y_c, y_b, y_t$. 
This integral has an extra $\rho'^{-2}$ factor compared to the $V_{\rm
UV}$ by the dimensional analysis. This makes the UV contributions more
important. 

Since the important integration region shifts to smaller $\rho'$ than
that for $V_{\rm UV}$, which we denote as $\rho$, we find
\beq
\laq{V2}
V_{\rm CPV}(a) \sim -\e \chi_0 
\left( \frac{C}{\rho^2 \L^2} \right)
\cos \left[ \frac{a}{f_a}+\theta_{\rm UV} \right] 
\eeq
with
\begin{align}
   C \sim \frac{\ab{C^{1111}_{ud}}}{y_uy_d}\sim 3\times 10^{9}
\ab{C^{1111}_{ud}},\quad
 \theta_{\rm UV}=\arg{C^{1111}_{ud}}, 
\end{align}
gives a conservative estimate of the CP violating effects. The
parameter $C$ can be very large but also can be of $O(1)$ if there are
chirality suppressions in the higher dimensional operator.
On the other hand, the phase $\theta_{\rm UV}$ is $O(1)$ in general.
Throughout this paper we assume $\theta_{\rm UV}=\O(1)$.

The experimental constraints from the neutron electric dipole moment
(EDM), $|d_n| \lesssim 3\times
10^{-26}e$\,cm~\cite{Baker:2006ts,Afach:2015sja}, put an upper bound
on the effective $\theta$ angle as $\ab{\bar{\theta}_{\rm
QCD}}\lesssim 2\times 10^{-10}$, where $\bar{\theta}_{\rm
QCD}=\vev{a/f_a}$, through the theoretical estimates of 
$
d_{n}= \(1.52 \pm 0.71\)\times 10^{-16} \bar{\theta}_{\rm QCD} e \,{\rm cm }~\AND~ d_{p}= \(-1.1\pm 1.0\)\times 10^{-16}\bar{\theta}_{\rm QCD} e \,{\rm cm },
$~\cite{Pospelov:2005pr, Dragos:2019oxn}.
Future storage ring experiments have sensitivities of $|d_{p}|\approx
10^{-29}$~\cite{Anastassopoulos:2015ura} which translates to
$|\bar{\theta}_{\rm QCD}|\sim 10^{-13}$.

\begin{figure}[!t]
\begin{center}  
   \includegraphics[width=105mm]{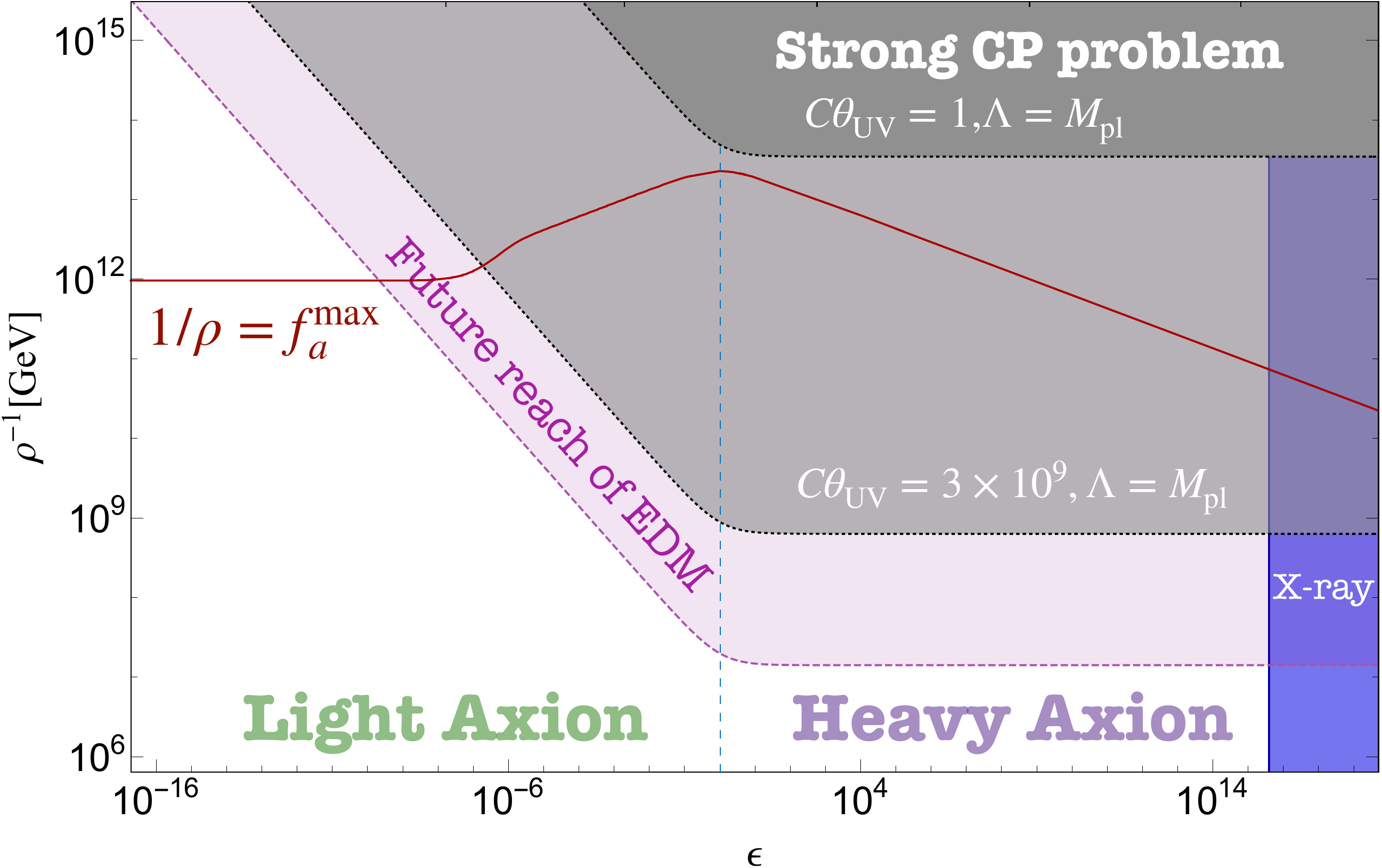}
      \end{center}
\caption{
The parameter region of the QCD axion with aligned small instanton in $\e\text{-}\r$ plane with $\L=M_{\rm pl}.$ 
The light gray and  gray regions are excluded by the too large neutron EDM with 
 $C \theta_{\rm UV}=3\times 10^9\AND 1,$ respectively. The purple shaded range above the dashed line may be tested in future EDM searches with $C \theta_{\rm UV}=3\times 10^9$. 
On the upper frame, the maximal axion decay constant from the axion abundance is denoted (see \Sec{window}). 
The red solid line represents $1/\rho= f_{a}^{\rm max}$. 
The blue region is excluded by  X-ray observation (by assuming dominant dark matter).
}\label{fig:Mpl2} 
\end{figure}

The condition for solving the strong CP problem can be obtained by calculating the VEV with the potential 
  \beq
  \laq{pot}
  V(a)= V_{\rm QCD}(a) +V_{\rm UV}(a)+ V_{\rm CPV}(a).
  \eeq
This results in
\begin{align}
-\vev{\frac{a}{f_a}}\sim C \theta_{\rm UV} \frac{\e}{1+\e} {(\rho \L)^{-2}}\lesssim  2\times  10^{-10}. 
\end{align}
In particular if $V_{\rm UV}$ dominates over the QCD potential, i.e.
in the heavy axion scenario, we need
\begin{align}
  \rho^{-1}\lesssim 3\times 10^{-10} \L 
  \left(
        {C \theta_{\rm UV} \over 3 \times 10^9 }
  \right)^{-1/2}.
\end{align}
For example, for $C \sim 3\times 10^9$ and $\Lambda \sim M_{\rm pl}$,
the instanton scale, $\rho^{-1}$, should be smaller than
$\O(10^9)\GEV$. 
 This condition does not depend on the size of the decay constant of the axion. 
The parameter region is shown in Fig.\ref{fig:Mpl2} in $\e-\rho^{-1}$ plane with $\L=M_{\rm pl}.$
The region above the light gray and gray range, respectively, with $C\theta_{\rm UV}=3\times 10^9\AND 1,$ are excluded due to the EDM bound. The purple region may be searched for in the future. 
The constraints and lines relevant to the axion dark matter will be discussed in \Sec{window}. 
From this figure, we can conclude that a heavy QCD axion, $\e>1$, must
accompanied with a large enough instanton size $\rho\gtrsim
10^{-10}\GEV^{-1}$ in the case of no chirality suppressions in the
higher dimensional operators, $C \sim 3 \times 10^9$. 
The constraints are milder for $C \sim 1$, but it is still important
to note that heavy axions cannot be realized by UV dynamics higher
than $10^{14}$~GeV. This fact is important in the discussion of the
axion in the extra-dimensional model or in general string axion
models.

\subsection{Relation to the quality problem of the PQ symmetry}

The CP violation we discussed is present even when the PQ symmetry is
exact (but anomalous), and thus known solutions to the quality problems may not work
for the UV instantons.

The discussion is qualitatively different depending on whether
$1/\rho$ is smaller or larger than the decay constant $f_a$.
In the case of $f_a\ll 1/\rho$, the discussion is similar to the
ordinary scenarios.
In this regime, one should consider the UV contributions to the PQ
breaking dynamics including the quark fields which make the PQ
symmetry anomalous.
For example, let us consider a model where a VEV of a scalar field
$\Phi_{\rm PQ}$ breaks the PQ symmetry spontaneously. In the presence
of a single pair of vector-like PQ quarks $Q_{\rm PQ}$ with an
interaction 
\beq 
\laq{PQ}{\cal L}\supset -y_\F
\F_{\rm PQ}\bar{Q}_{\rm PQ}Q_{\rm PQ}
\eeq 
the potential for $\Phi_{\rm PQ}$ receives contributions from small
instantons which scale as $\propto y_\F \F_{\rm PQ} \rho^{-3}$ or $y_\F
\F_{\rm PQ} \rho^{-5}\L^{-2}.$ 
The axion appears as the pseudo Nambu-Goldstone boson associated with
the PQ breaking, $\langle \Phi_{\rm PQ} \rangle \sim f_a \ll 1 / \rho$. The linear
term of $\Phi_{\rm PQ}$ from the UV instanton contributes to the axion mass.

The appearance of the linear term can be avoided if $\F_{\rm PQ}$ is
charged under some gauge symmetry like gauged $Z_N$ symmetry (and also
we need non-trivial PQ quark contents to have color anomaly-free
interaction), as in the solutions to the ordinary quality
problem~\cite{Chun:1992bn,BasteroGil:1997vn,Babu:2002ic}\cite{Fukuda:2017ylt,Duerr:2017amf,Bonnefoy:2018ibr}\cite{Randall:1992ut,DiLuzio:2017tjx,Lillard:2018fdt, Lee:2018yak, Ardu:2020qmo, Yin:2020dfn, Yamada:2021uze}. 
By symmetry, the small instanton contribution is suppressed as
$\propto \F_{\rm PQ}^N \rho^{N-4} \OR \F_{\rm PQ}^N \rho^{N-6}
\L^{-2}$, and the problem may be solved with large enough $N$. At the
same time we also have the suppressed  aligned contribution.

For $\vev{\F_{\rm PQ}} \sim f_a \gtrsim 1/\rho$, the problem is more
serious. The reason is that the PQ field $\F_{\rm PQ}$ appears in the
instanton through the $f_a$ dependence of the running gauge coupling,
which means the dependence is a singular $\log{\F_{\rm PQ}^N}$ with
the $Z_N$ gauge symmetry. This is no longer suppressed by the scales
of higher dimensional terms with larger dimension for large $N$. The
approach by the $Z_N$ symmetry does not work in this case.

In appendix~\ref{chap:model}, a heavy axion model to avoid the UV
problem is discussed.

Although we focused on the new CP violation in the axion models, there
should be a similar CP violating effect in the case of the massless up
quark solution discussed in Refs.~\cite{Agrawal:2017evu,Gupta:2020vxb}. If we take into account various CP-violating higher
dimensional terms, the up quark mass would receive an additional phase
of $\propto (\rho \L)^{-2}$.  
Therefore to generate the mass and preserve the vanishing strong CP
phase, we need the instanton size to be large enough. One needs to
check if the enough size of up-quark mass at low energy is realized in
this case.

\section{Axion dark matter with UV instantons}
\lac{window}

In this section, we study the cosmological abundance of the QCD axion
by taking account of the small instanton.

The decay constant of the QCD axion has a window in which the axion
can have a consistent cosmological history (without fine-tuning):
\beq
10^8\GEV \lesssim f_a\lesssim f_a^{\rm max}\simeq 10^{12}\GEV.
\eeq
The lower bound comes from the duration of the neutrino burst in
SN1987A~\cite{Mayle:1987as,Raffelt:1987yt,Turner:1987by,Chang:2018rso}, the upper bound, $f_a^{\rm max}$, is obtained from the
over production of the QCD axion in the early Universe, i.e. the axion
dark matter is over abundant. The axion dark matter is realized around
$f_a\sim f_a^{\rm max}$ with the initial misalignment angle
$|\theta_i|\sim 1.$

The estimation of the abundance relies on the temperature dependence
of the topological susceptibility~\cite{Abbott:1982af,
Preskill:1982cy, Dine:1982ah}. (See Refs.~\cite{Berkowitz:2015aua, Kitano:2015fla, Borsanyi:2015cka, Frison:2016vuc, Borsanyi:2016ksw}
for recent lattice computations.)  When the QCD axion mass becomes
comparable to the Hubble parameter, the axion starts to oscillate
around the potential minimum from its initial misalignment angle
$\theta_i$. The axion number conserves and later the axion energy
density behaves as dark matter. 
The temperature dependence of the axion mass determines when the axion starts to oscillate. The inclusion of the aligned small instanton contribution changes this
temperature dependence, and the estimation of the abundance becomes non-trivial. In the following, we will estimate carefully the
abundance of the axion in the presence of the small instanton.

We assume $\bar{\theta}_{\rm CP} \ll1$ and neglect the contribution of
CP violating term for a while: i.e. we use the approximation of
\beq
V\approx V_{\rm QCD}+V_{\rm UV}.
\eeq
Then, the axion physical mass at the vacuum can be obtained as
\beq
m_{a}^2[\e]= \frac{\chi_0(1+\e) }{f_a^2}=m_{a}^2(0)(1+\e).
\eeq
The abundance formula is known for $\e \gg 1$ or $\ll 1$ since we can neglect 
$V_{\rm QCD}$ or $V_{\rm UV}$. 
With only $V_{\rm UV}$ we obtain~(we use the fit in Ref.~\cite{Ho:2019ayl})
\beq
\Omega_{a} h^2 \approx 0.05 \e^{1/4}\,
\bigg(\frac{g_{\star,\text{osc}}}{18}\bigg)^{-1/4}
 \(\frac{\theta_i}{0.001}\)^2\bigg(\frac{f_a}{10^{17}\,\text{GeV}}\bigg)^{3/2} \text{    if }\e \gtrsim 1.\laq{absa} 
\eeq
Here $g_{\star,\text{osc}}$ is the relativistic degree of freedom at the onset of oscillation. 

With only $V_{\rm QCD}$ we obtain~\cite{Ballesteros:2016xej}
\begin{eqnarray}
\laq{QCDaxion}
\Omega^{}_{a} h^2 
\,\approx\, 
0.35 \left(\frac{\theta_i}{0.001}\right)^{2}\times
\begin{cases}
\displaystyle 
\left(\frac{f_a}{3\times 10^{17}\,{\rm GeV}}\right)^{1.17} 
& f_a \,\lesssim\, 3 \times 10^{17}\,{\rm GeV} \vspace{3mm}\\
\displaystyle 
\left(
\frac{f_a}{3\times 10^{17}\,{\rm GeV}}\right)^{1.54}
& f_a \,\gtrsim\, 3 \times 10^{17}\,{\rm GeV}
\end{cases} \text{    if }0<\e \ll 1.
\end{eqnarray} 
The difference of the two cases comes from the temperature dependence of the potential.

For $\e\sim 1$ (more precisely $10^{-13}\text{-}10^{-6}\lesssim \e\lesssim 1$ for $f_a=10^{8}\text{-}10^{14}\GEV$)
, on the other hand, the abundance gets suppressed compared with $\e \ll1$ because in the presence of both terms, the time for the onset of oscillation is faster than the case of QCD potential only.

To check this we solve
\beq
\ddot{a}+3H \dot{a} =-\frac{\partial V}{\partial a},\ebq \dot{s}+3H s=0 
\eeq
with $H=\sqrt{\frac{\pi^2 g_{\star}}{30} \frac{T^4}{M_{\rm pl}^2}} $ being the Hubble parameter
in the radiation dominated Universe, $s=\frac{ 2\pi^2 g_{\star s}}{45} T^3$  the entropy density, and $g_\star$ ($g_{S\star}$) the relativistic degrees of freedom (for entropy) at the cosmic temperature $T$. 
We evaluate the abundance parameter defined by 
\beq
\Omega_a[t] \equiv \frac{\rho_a}{s} \frac{s_0}{\rho_{\rm crit}} 
\eeq 
with $\rho_a[t]=\frac{\dot{a}^2}{2}+ V(a),$ and $s_0$ and $\rho_{\rm
crit}$ the current entropy density and critical density, respectively.
When $T\ll T_{\rm c}$, we can easily show that this value conserves.
We should also compare this with the observed dark matter abundance
$\Omega_{\rm DM}h^2\approx 0.12$~\cite{Aghanim:2018eyx}. 
\begin{figure}[!t]
\begin{center}  
   \includegraphics[width=105mm]{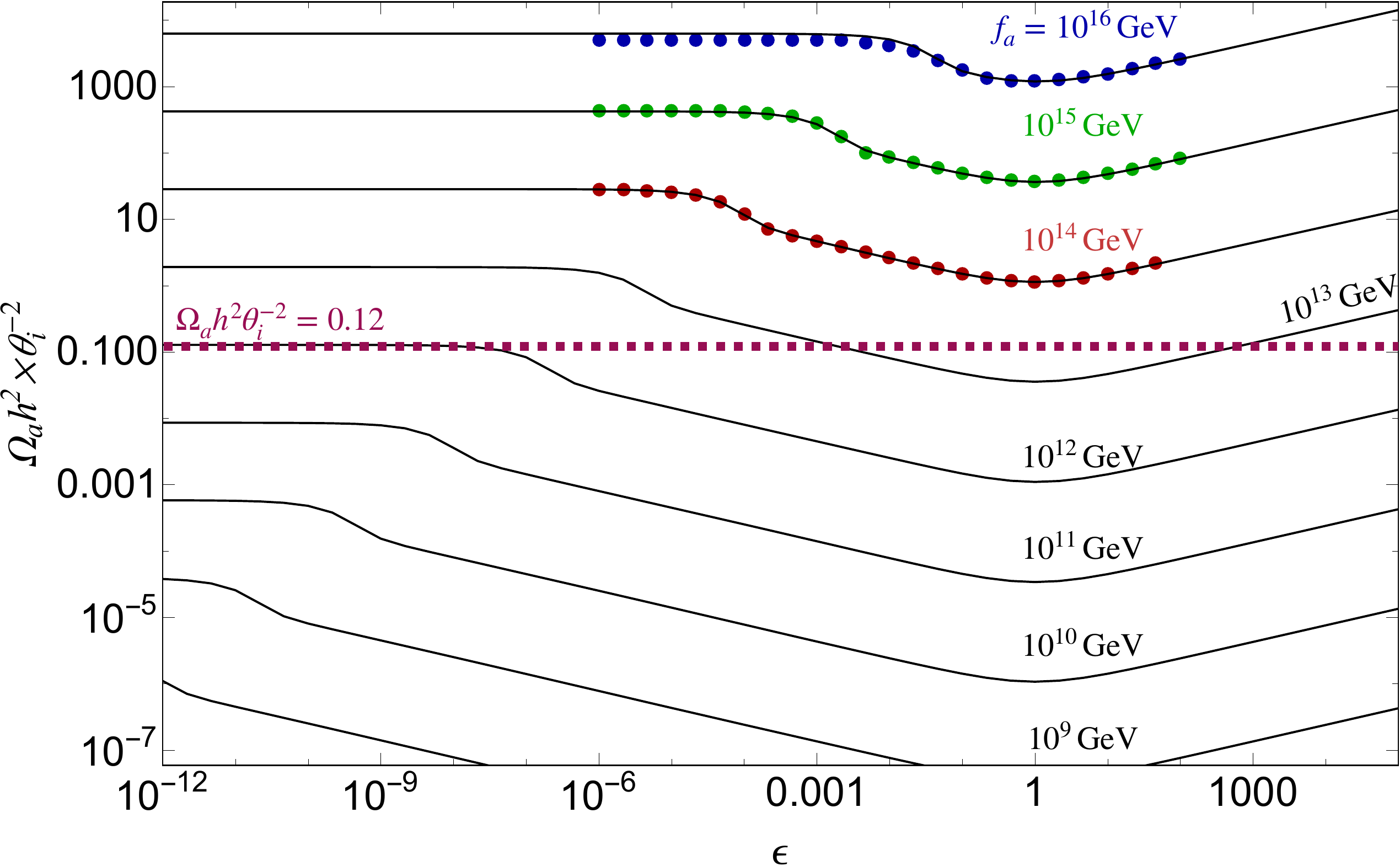}
      \end{center}
\caption{
The resulting abundance divided by $\h_i^2$ for the QCD axion with aligned small instanton by varying the relative instanton contribution, $\e$. 
The data points are calculated numerically by solving the equation of motion and are for  $f_a=10^{16,15,14}\GEV$ from top to bottom.  The solid lines are fitted from the data and correspond to $f_a=10^{16,15,14,13,12,11,10,9,8}\GEV$ from top to bottom. The correct dark matter abundance with $\theta_i=1$ is denoted by the purple dotted line. 
}\label{fig:DM} 
\end{figure}
The numerical result (where we approximate the potential by the quadratic term) by varying $\e$ is shown in Fig.\,\ref{fig:DM} for the blue, green and red points, respectively, with $f_a=10^{16,15,14}\GEV$ from top to bottom.

For later convenience, we find a fitting formula for the abundance 
\beq
\Omega_a [t\to \infty]\to \left. C_{\rm fit,1}\sqrt{\frac{\chi_0 (1+\e)}{f^2_a}}\times \frac{ \sqrt{(C^{\rm fit,2}\chi(T)+\e \chi_0)/f_a^2} \theta_i^2 f_a^2 }{s} \frac{s_0}{\rho_c}\right|_{T=T_{\rm fit,osc}}
\eeq
where $T_{\rm fit,osc}$ is obtained by equating $m^{\rm fit}=H$, with
$m^{\rm fit}\equiv \sqrt{(C^{\rm fit,3}\chi(T)+\e \chi_0)/f_a^2}$.
Here $C^{\rm fit,i}\approx\{1.03, 46.2, 5.35\}$ is obtained by fitting
the numerical data (red points in Fig.~\ref{fig:DM}) at
$f_a=10^{14}\GEV.$ This formula agrees well  also with the numerical
results for $f_a=10^{15,16}\GEV$. We display the abundance in
Fig.~\ref{fig:DM} for analytical results of  smaller $f_a$, which
requires calculation costs if we solve the equation of motion.
Importantly,  $10^{-13}\lesssim \e\lesssim 1$, the abundance can be
affected by the aligned instanton contribution, which can suppress the
abundance by $\O(10^{-3}-1)$ for the axion with $f_a\gtrsim 10^8\GEV$.
This suppression opens the axion window up to 
$ f_a^{\rm max}[\e]\lesssim 10^{13\text{-}14}\GEV$
with the aligned small instanton. 

We overlay the maximal value of the decay constant allowed by the axion abundance as a function of $\e$ in Fig.~\ref{fig:Mpl2} (red solid line). We took $\theta_i=1$ in the figure. 
The constraints
form X-ray observation are translated from Ref.\,\cite{Essig:2013goa}, by
assuming dominant axion dark matter (whose photon coupling is induced through the meson-axion mixing) from the
misalignment mechanism in blue shaded region. The mass corresponds to
$m_a \sim 5\KEV$ around the boundary.\footnote{The axion dark matter
may also be produced from inflaton decay~\cite{Moroi:2020has}. In this
case to suppress the misalignment contribution we need to have smaller
$f_a$. Then the bound moves towards a smaller value of $\e$.  }
The upper bound of the axion window or the natural decay
constant for the axion dark matter is \beq 10^{10}\GEV \lesssim
f_a^{\rm max}[\e]\lesssim 10^{14}\GEV \eeq by varying $\e$.

In concrete models, $\epsilon$, $\rho^{-1}$ and $f_a^{\rm max}$ are
related.
For example, in the model of Appendix~\ref{chap:model}, $\rho^{-1}\sim f_a$.
In this model, the heavy axion ($\e>1$) is difficult to be the
dominant dark matter from misalignment otherwise the strong CP problem
exists unless $C$ has chirality suppressions.  For heavy axion dark matter where we
should have $f_a\sim 1/\rho \lesssim 10^{9}\GEV$ we need to consider
other mechanisms to produce the correct abundance~e.g.
\cite{Daido:2017wwb, Co:2017mop, Co:2018mho, Takahashi:2019pqf,Takahashi:2019qmh, Kobayashi:2019eyg, Nakagawa:2020eeg, Huang:2020etx,Moroi:2020has, Nakagawa:2020zjr }. This scenario needs to evade the
SN1987A constraint, $f_a \sim 1/\rho \gtrsim 10^8\GEV$, and so the
heavy axion dark matter is likely to be fully tested in the future EDM
experiment.

Before ending this section, let us comment on $-1<\e< 0$.  In this
case the axion starts to oscillate towards $a/f_a\approx \pi$ in the
early Universe and when IR contribution dominates the oscillation is
around the $a\simeq 0.$ We do not consider this possibility because
from phenomenological side it will cause a serious domain wall
problem.  If $\e<-1$, it predicts $\theta_{\rm CP}=-\pi$, which is
excluded from observation. 
In a wide class of models, we expect that the sign of the UV
contributions is the same as the IR one since it is ensured by the
positive path-integral measure of QCD~\cite{Vafa:1984xg}.

\section{Small instantons in the extra-dimension scenarios} 

\lac{ext2}
In certain models, the integral \eq{vac} is dominated at around the UV
cutoff, i.e. $\rho\sim 1/\L_{\rm cutoff}.$ Once this is the case, the
CP violation contribution \eq{CPV} is also UV dominated, which gives
\beq
\laq{CPV2}
V_{\rm CPV}\sim  - \e \chi_0 C \frac{\L_{\rm cutoff}^2}{\L^2}
\cos \left[ \frac{a}{f_a} +\theta_{\rm UV} \right],
\eeq
by assuming $\L_{\rm cutoff}\ll \L.$

Let us first consider the case where the higher dimensional operators
have no chirality suppression. In this case, $C\sim 3\times 10^{9}$
gives a large enhancement. For $\L_{\rm cutoff}/\L \gtrsim y_i$, the
dominant contribution is the diagrams with multiple insertion of the
higher dimensional operators. In particular with $\L_{\rm cutoff}\sim
\L$, we obtain
\beq
\laq{superUV}
V_{\rm CPV}\sim  - \e \chi_0 C \frac{\L_{\rm cutoff}^6}{\L^6}
\cos \left[\frac{a}{f_a} +\theta_{\rm UV} \right].
\eeq
Now $C\sim 1/(y_u y_d y_s y_c y_t y_b)\sim 10^{18}$ and $\theta_{\rm
UV}=\O(1).$
To evade the EDM constraint we need 
\beq
\e C \theta_{\rm UV} \frac{\L_{\rm cutoff}^6}{\L^6}\lesssim 10^{-28}.
\eeq

To be more concrete, let us study the model where the axion is
originated from the gauge field in the extra dimension. The model has
a nice feature to protect the PQ symmetry by the gauge invariance and
the locality, and catches the essential feature of axions in string
theories.

A simple realization is possible in five dimensional model with the
fifth dimension compactified by an orbifold
$S_1/Z_2$~\cite{Choi:1996fs}.  The gluon lives in the fifth
dimensional bulk with size $R$ and the axion is identified as the
Wilson line of the fifth direction of a $\U(1)$ gauge field. 
The cut-off scale of the theory is identified as the 5d planck scale,
 \beq \L_{\rm cutoff} \sim  (M_{\rm pl}^2/R)^{1/3}, \eeq if there is
no other interactions which gets strong below this scale. 
We consider the case where the quarks and the Higgs field live on the
orbifold fixed point. 
The PQ symmetry is identified as the shift symmetry of the $A_5$
component of the $\U(1)_{\rm PQ}$ gauge field on the boundary. While it is
protected by the gauge symmetry and locality, the mixed Chern-Simons
term,
\begin{align}
   {k \over 8 \pi^2} \int A_{\rm U(1)_{\rm PQ}} F_{\rm SU(3)_c} F_{\rm SU(3)_c},
\end{align}
can give the anomalous coupling between the axion and the gluons.
This setup provides a very simple origin of the axion (i.e. string
axion).
In this model, the axion has a decay constant 
\beq
f_a \sim 1/R.
\eeq

On the brane, we have ``Planck scale" suppressed terms similar to
\eq{dim6}\footnote{ If in a setup $u$ and $d$ are sequestered,
\eq{dim6} may be suppressed. In this case, we can consider $ H
\bar{Q}u |H|^2$. This term should exist as long as the Yukawa coupling
exist on the brane.} i.e. 
\beq \L\sim \L_{\rm cutoff}.\eeq
The aligned contributions is studied in Ref.~\cite{Gherghetta:2020keg}, where the 5d instanton solution is found to contribute as
\beq
\laq{aligned}
\chi_{\rm aligned}[R]= \e[R] \chi_0 =  \Pi_{i}\(\frac{y_i}{ 4\pi}\) \int_{1/\L_{\rm cutoff}}^{R}{\frac{d \rho' }{\rho'^5} e^{-S_{\rm eff}[1/\rho']}  F^{(\rm 5D)}[\rho']}.
\eeq
where
\beq
\laq{Seffextra}
F^{(\rm 5D)}[\rho']
= 1.5\times 10^{-3}\(\frac{2\pi}{\a_s(1/R)}\)^6, S_{\rm eff}= \( \frac{2\pi}{\a^{\rm SM}_s(1/R)}- 3\xi(R/\rho') \frac{R}{\rho'}+
b_0 \log\(\frac{R}{\rho'}\)\). 
\eeq
The function, $\xi$, which increases to approach to $\sim 0.35$ with
$R/\rho'\to R \L_{\rm cutoff}$, can be found in
\cite{Gherghetta:2020keg}. The beta function coefficient, $b_0$, and the coupling constant, $\a_s^{\rm SM},$ are, respectively, given by 
\beq b_0=7 \AND 1/\a^{\rm SM}_s(1/R)\equiv 1/\a_{s,{\rm EW}}-b_0/(2\pi)
\log{(v_{\rm EW} R)}\eeq with $\a_{s,{\rm EW}}=0.118$ and $v_{\rm
EW}=100\GEV.$ The suppression of $S_{\rm eff}$ at large $1/\rho'$ makes
the integral dominated around $\rho'\sim 1/\L_{\rm cutoff}.$ 
Here, the fundamental 5d gauge coupling, $g_5^2$, (dimension -1) is matched to the SM gauge coupling as
\beq
\laq{g5}
g_5^2= 4\pi^2 R \times \alpha^{\rm SM}(1/R).
\eeq 
Since the gauge interaction is from a higher dimensional term, the theory gets strong at a scale $\sim 24 \pi^3/g_5^2$.

Considering a diagram as in Fig.~\ref{fig:2}, we obtain
\begin{align}
\laq{CPV2}
\chi_{\rm CPV}[R]
\sim 
\(\frac{1}{\L^2}\) (y_uy_d)^{-1}\(\Pi_{i}\frac{y_i}{4\pi}\)\int_{1/\L_{\rm cutoff}}^{R}{\frac{d \rho' }{\rho'^5}  \rho'^{-2} e^{-S_{\rm eff}[1/\rho']}  F^{(\rm 5D)}[\rho']}.
\end{align}
By numerically performing this integral, we confirm the previous general argument at a good precision with $C \sim 10^{9}$ by taking the ratio of \Eq{CPV2} to \Eq{aligned} with $\L\sim \L_{\rm cutoff}.$
In this model, however, the dominant contribution is from the term corresponding to \eq{superUV}. 
\beq
\chi_{\rm CPV}[R]\sim  
\(\frac{1}{16\pi^2 \L^2}\)^3\int_{1/\L_{\rm cutoff}}^{R}{\frac{d \rho' }{\rho'^5}  \rho'^{-6} e^{-S_{\rm eff}[1/\rho']}  F^{(\rm 5D)}[\rho']}.
\eeq
This is again consistent with our general argument.

In the case of $\L\sim \L_{\rm cutoff}$, however, it is natural to
assume that there are chirality suppressions for the higher
dimensional terms, so that the small coupling constants, such as $y_u$
and $y_d$, are stable under the radiative corrections.
Then \eq{dim6} is in the form,
\beq
C_{ud}^{ijkl}=C_{ud} Y_u^{ij} Y_d^{kl},
\eeq
with
\beq
C_{ud}\sim \O(1)+i \O(1).
\eeq
Then we obtain 
\beq
\laq{MFVCPV}
\chi^{\rm CS}_{\rm CPV}[R]= \(\frac{1}{\L_{\rm eff}^2}\)\(\Pi_i\frac{y_i}{4\pi}\) \int_{1/\L_{\rm cutoff}}^{R}{\frac{d \rho' }{\rho'^5}  \rho'^{-2} e^{-S_{\rm eff}[1/\rho']}  F^{(\rm 5D)}[\rho']}.
\eeq
Here we introduced $\L_{\rm eff}$ for a later convenience. This
satisfies $\L_{\rm eff}\sim \L \gtrsim \L_{\rm cutoff}$ and it
includes $|C_{ud}|$ as well as relative numerical uncertainty
compared with the integral in \Eq{aligned}. This contribution may be
suppressed compared with the aligned instanton constribution by
$\(\L_{\rm cutoff}/\L_{\rm eff}\)^2.$

\begin{figure}[!t]
\begin{center}  
      \includegraphics[width=105mm]{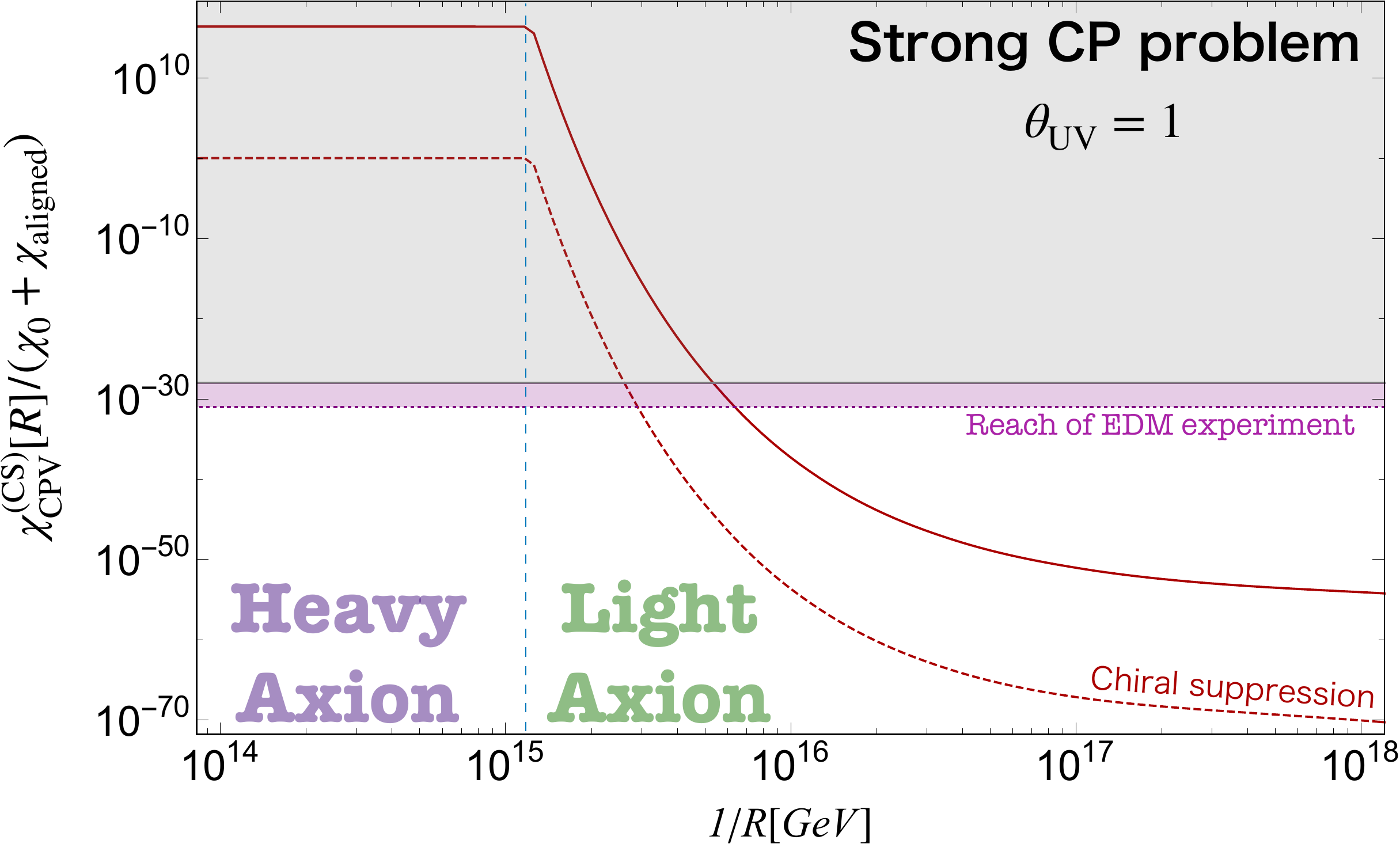}
      \end{center}
\caption{Allowed regions for the extra-dimension model by numerically
integrating the small instanton contributions \eq{aligned},  \eq{CPV2} and \eq{MFVCPV}, 
with $\L^3R= M_{\rm pl}^2$ and $\L=\L_{\rm cutoff}$. 
The red solid (dashed) line represents $\chi_{\rm CPV}/(\chi_{0}+\chi_{\rm aligned})$ ($\chi_{\rm CPV}^{\rm CS}/(\chi_{0}+\chi_{\rm aligned})$).
The EDM bound (gray
region) and the future reach (purple region) is estimated by assuming
$\theta_{\rm UV}\sim 1.$ }\label{fig:extraCPV} 
\end{figure}

The resulting parameter region by calculating $\chi_{\rm
 CPV}/(\chi_{0}+\chi_{\rm aligned})$ is shown in
 Fig.~\ref{fig:extraCPV} by the red solid line. The red dashed line in
 Fig. \ref{fig:extraCPV} is for the chirality-suppressed case by
 assuming $\L_{\rm eff}=\L=\L_{\rm cutoff}.$ The gray (purple) region
 is the bound (future reach) from nucleon EDM experiments by assuming
 $\theta_{\rm UV}\sim 1$. 
 Interestingly with $1/R\gtrsim 10^{15-16}\GEV$ we have an allowed
region. This is because in this setup $g_5^2 \L_{\rm cutoff} \propto
(M_{\rm pl}^2 R^2)^{1/3}\a^{\rm SM}_s(1/R)$ which makes the gauge
interaction at around the cut-off weaker for smaller $R$. In this
regime, the axion mass is dominated by the IR QCD contributions, and
the low-energy axion physics is the same as the conventional one.

\section{Cosmology of extra-dimensional axions}

The scenario of the axion from extra-dimension is a 
 natural possibility for the solution of the strong CP problem. We
have seen in the previous section that the radius of the extra
dimension is required to be above the scale of the grand unification theory (GUT),
$10^{15}~{\rm GeV}$, in order to avoid yet another strong CP problem
from the UV instantons. The axion decay constant $f_a$ is also
required to be above the GUT scale. In other words, we need the gauge
coupling to be weakly coupled up to the 5d Planck scale also for
solving the strong CP problem.
This sounds quite natural in string theories or in general in the
theory of quantum gravity, where the typical scale is the GUT or the
Planck scale. However, cosmologically, it implies that the axion has
an overproduction problem if we do not require a fine-tuning of the
initial misalignment angle $\theta_i.$

In the extra-dimensional scenario, we find that there is a simple
possibility to avoid the overproduction and realize the axion dark
matter naturally by the help of the UV instantons.
We mainly consider the case where the higher dimensional operators
have chirality suppressions. The case with no  chirality suppression
will also be discussed later.

\subsection{Natural axion dark matter}

In the extra dimensional scenario, there is a modulus field, the
radion, to represent the size of the extra dimension.
In the current Universe, the radion field $r$ is stabilized at
$\vev{r}=R$. 
By saying that the radion is a dynamical field, we implicitly assume that the mass, $m_r^2\equiv V''[r]$, is not extremely heavy. We will come back to the dynamics of the radion in Sec.\,\ref{sec:radion}

Then, it is possible during the inflation that $r$ is
displaced from the current location. This changes the coupling of 4d
gravity via
\beq
{\cal L}\sim \L^3 r {\cal R}. 
\eeq
In the following we perform a Weyl transformation to move to the
Einstein frame, so that the coefficient of the Ricci scalar is always
$M_{\rm pl}^2.$ Instead, various potentials at the zero temperature
obtain an additional factor of ${M_{\rm pl}^4}/{(\L^3 r)^2}.$ 
For
instance the axion potential from QCD is
\beq
\laq{QCDpot2}
\tl{V}_{\rm QCD}[r,a]=\tl{\chi}_{\rm QCD}[r]
\(1-\cos{\frac{a}{f_a}}\).
\eeq
Here we use 
\beq
\laq{QCDr}
\tl{\chi}_{\rm QCD} [r] \sim \frac{M_{\rm pl}^4}{(\L^3 r)^2}  (\Pi_i y_i ) \L_{\rm QCD}[r]^4
\eeq
 with 
\beq
\L_{\rm QCD}[r]\approx \exp{\(-\frac{2\pi}{b_0\a_s(1/r)}+\frac{2\pi}{b_0\a^{\rm SM}_s(1/R)}\)} \L_{\rm QCD}[R]
\eeq
and $\L_{\rm QCD}[R]\sim 400\MEV.$ 
We emphasize that 
\beq
\a_s[1/r]=\frac{g_5^2}{4\pi^2 r }
\eeq
and  $g_5^2$ is fixed by \eq{g5} once $\vev{r}=R$ is given. 
This form is a good approximation when
the QCD scale, $\L_{\rm QCD}[r]$, is higher than the electroweak scale.
The chirality flips are supplied by the Yukawa interactions whereas
the VEV of the Higgs field is induced by the chiral symmetry breaking
of $\SU(6)\times \SU(6)\to \SU(6)$. Due to the top quark condensation,
the Higgs field obtains an expectation value $\vev{H}_r \sim \L_{\rm
QCD}[r],$ with $\vev{}_r$ denotes the expectation value in a spatially
homogeneous radion background $r$. 
More precisely, there could be CP-violating effect to the IR QCD contribution via not only the higher dimensional operators but also from the SM interactions through the CKM phase~e.g. \cite{Pospelov:2005pr}. The latter contribution is estimated to be at most comparable to the QCD potential in \Eq{QCDpot2}. Although this could be important in some cases, we are particularly interested in the case that $\tl{V}_{\rm QCD}$ is not the dominant contribution and thus the effects can be ignored in the most successful parameter region.  One should keep in mind that there can be significant corrections from weak interactions when  $\tl{V}_{\rm QCD}$ is important.

 \begin{figure}[!t]
\begin{center}  
      \includegraphics[width=105mm]{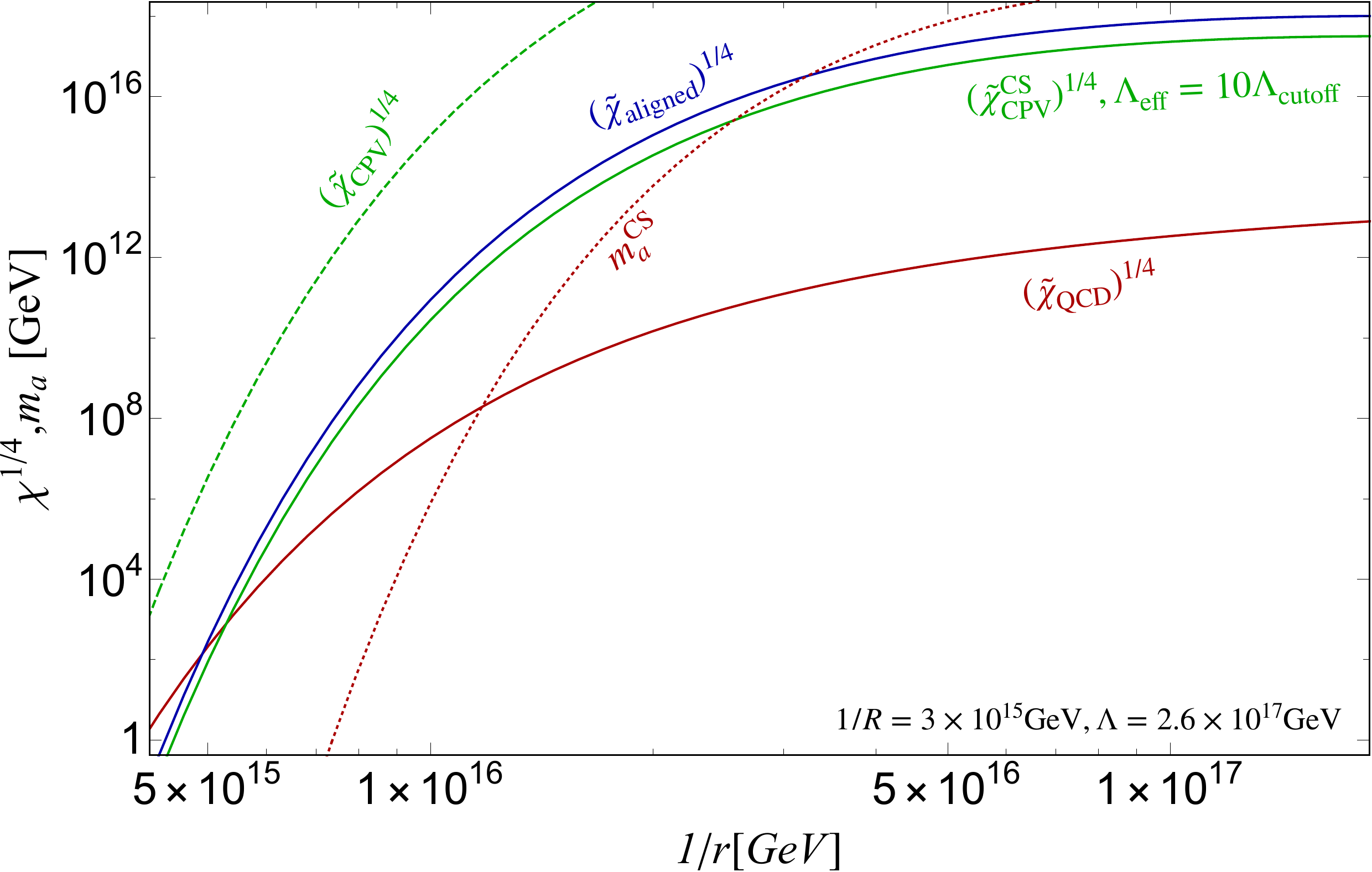}
      \end{center}
\caption{
Various contributions and axion mass (with chiral suppressions) with a dynamical radion field, $r$. We fix $1/R=3\times 10^{15}\GEV, \L=\L_{\rm cutoff}=1/10 \L_{\rm eff} \AND \L^3 R=M_{\rm pl}^2.$}\label{fig:CPV} 
\end{figure}

On the contrary the CP violating small instanton contribution is estimated as 
\beq
\tl{V}_{\rm CPV}[r,a]\approx -\tl{\chi}^{\rm (CS)}_{\rm CPV}[r]\cos{[\frac{a}{f_a}+\theta_{\rm UV}]}. 
\eeq
Here, 
\beq 
\tl{\chi}^{\rm (CS)}_{\rm CPV}[r] =\frac{M_{\rm pl}^4}{(\L^3
r)^2}\chi^{\rm (CS)}_{\rm CPV}[r]. \eeq
Similarly we get the aligned contribution, 
\beq
\tl{\chi}_{\rm aligned}[r]=\frac{M_{\rm pl}^4}{(\L^3 r)^2} \chi_{\rm aligned}[r].
\eeq
$\tl{\chi}^{\rm (CS)}_{\rm CPV}[r], \tl{\chi}_{\rm QCD}[r] \AND \tl{\chi}_{\rm aligned}[r] $ as well as $m_a^{\rm CS}[r]\equiv  r\sqrt{\tl{\chi}_{\rm QCD}[r]+\tl{\chi}_{\rm aligned}[r]} $\footnote{We add a superscript, CS, to denote that this form is justified in the chiral suppression scenario. Otherwise $\tl{\chi}_{\rm CPV}$ may be dominant. } are evaluated numerically and are shown in Fig.~\ref{fig:CPV}. 
We fix $1/R=3\times 10^{15}\GEV$.
First, we can see that the
all contributions increase with increasing $1/r$. This is because the coupling $\a_{s}[1/r]\propto 1/r$ increases. Thus the exponential suppressions in all instanton contributions are alleviated (see  Refs.\,\cite{Dvali:1995ce,Banks:1996ea,Choi:1996fs,Jeong:2013xta,Co:2018mho,Matsui:2020wfx} for stronger QCD, Ref.\,\cite{Buen-Abad:2019uoc} for stronger small instanton by enhancing Yukawa couplings.).
Second, the CPV contribution as well as the aligned contribution increases faster than the IR QCD contribution does. This is because the exponential suppression in $\chi_{\rm aligned} \OR \chi^{\rm (CS)}_{\rm CPV}$,  $\propto e^{-\frac{2\pi}{\a_s[1/r]}}$ is larger than that of $\chi_{\rm QCD}$, $\propto e^{- \frac{4}{7}\times\frac{2\pi}{\a_s[1/r]}} .$ (Here we neglect the second and third terms in \eq{Seffextra} for analytic argument since they are irrelevant in the weakly coupled region to satisfy the EDM bound.) 
At larger $1/r$, the former increases faster. 

This feature is important in our mechanism: 
 if $r=r_{\rm inf}$ during inflation is smaller than $R$, 
one may obtain the UV instanton contribution dominating over the IR contribution. 
Furthermore, the axion mass can easily satisfy
\beq 
\laq{condmech}m_a[r_{\rm inf}]\gtrsim H_{\rm inf}\eeq
with $H_{\rm inf}$ being the Hubble parameter during inflation.  
We find that the axion, then, is stabilized at a CP-violating position, 
\beq
\theta_{r_{\rm inf}}\sim  \theta_{\rm UV}\min[ \chi_{\rm CPV}^{(\rm CS)}[r_{\rm inf}]/\chi_{\rm aligned}[r_{\rm inf}],1].
\eeq 
After the inflation the radion either rapidly or slowly settle into the minimum of the potential~\cite{Linde:1996cx,Nakayama:2011wqa}.
In any case, since the IR QCD contribution is absent due to the finite temperature effect from inflaton decays (by assuming the SM radiation temperature is higher than the QCD scale), 
$\theta_{r_{\rm inf}}$ is kept intact after inflation until the temperature decreases to around the QCD scale. 
This means
\beq
\theta_i= \theta_{r_{\rm inf}}.
\eeq 
If the chiral suppression were absent in the higher dimensional
operators, this mechanism would give the initial misalignment of order
$\theta_{\rm UV}\sim 1$, and the axion would be overproduced.

On the other hand, with chirality suppressions, we find
\beq
\theta_i \sim \theta_{\rm UV} \frac{\tl{\chi}^{\rm CS}_{\rm
CPV}}{\tl{\chi}_{\rm aligned}}\sim \theta_{\rm UV}  \(\frac{ \L_{\rm
cutoff}}{\L_{\rm eff}}\)^2 \sim  1~\% \times \theta_{\rm UV} \(\frac{\L_{\rm
cutoff}/\L_{\rm eff}}{0.1}\)^2.
\eeq
Consequently the axion dark matter can be explained by using \Eq{QCDaxion} in a wide range of
parameters satisfying $H_{\rm inf}\lesssim m_a[r_{\rm inf}]$ 
and UV instanton dominance. 
Although we need a mild
tuning on $\L_{\rm cutoff}/\L_{\rm eff}$, 
 $\L_{\rm
cutoff}/\L_{\rm eff}<1$ may be needed for a weakly coupled theory. 

The allowed region in Fig.~\ref{fig:CPV} is consistent with the upper
bound of 
\beq 
\laq{Hinfup}
H_{\rm inf}\lesssim 6\times 10^{13}\GEV
\eeq 
from the constraint for the tensor-to-scalar
ratio~\cite{Akrami:2018odb} for $1/r_{\rm inf}\gtrsim 10^{16}\GEV$.
On the other hand, in the region $1/r_{\rm inf}\gg 10^{17}\GEV$,
should be avoided since the potential energy from QCD exceeds that
during the inflation.

\subsection{Dynamics of $r$ and parameter regions}

\label{sec:radion}

Let us consider the radion mass during inflation. We may take the
radion nearly massless in 5d by assuming that the brane position is
not strongly fixed. Since the radion kinetic term in the Einstein
frame is given by (e.g.~\cite{Ponton:2001hq})
\beq
{\cal L}= \frac{3}{4} M_{\rm pl}^2 (\partial \log{r} )^2,
\eeq
it couples to other fields through Planck suppressed operator with normalized kinetic terms. 
 In any case there is a radiative induced mass after the compactification. 
 This mass may be dominated by the bulk gauge/gravity
 interaction~\cite{Ponton:2001hq} $\sim {1 / (16\pi^2R^4 M_{\rm pl}^2)}$,
 which sets the lower bound of the radion mass without a tuning:
\beq
\laq{RLB}
m_r[R]\gtrsim 10^{10}\GEV \(\frac{10^{15}\GEV}{1/R}\)^2.
\eeq
In addition the brane localized potential may give heavier mass to the radion.
In the following we take  $m_r^2$ arbitrary to check which mass range
of the radion is consistent with the previous discussion. 
While we implicitly assume that the inflaton field to drive inflation
is not the radion itself, it is possible to identify those two fields
as we briefly mention in \Sec{unif}.

During the inflation, the radion field value may be different from $R$. In general the radion would acquire a run-away potential $
V_{\rm runaway}\sim \(\frac{M_{\rm pl}^2}{r \L^3 }\)^2 H_{\rm inf}^2 M_{\rm pl}^2
$. This biases $r$ to a larger value. The localized radion potential on the brane may involve both inflaton and radion.  This can give an interaction between the radion and inflaton. 
We note that there is also a contribution from the QCD instanton to the radion mass squared $\d m_r^2[r]\sim \frac{\sqrt{\tl{\chi}_{\rm aligned}+ \tl{
\chi}_{\rm QCD}}}{M_{\rm pl}}$ which is always subdominant to the Hubble induced mass if the QCD potential is negligible compared to inflation energy density.

To be generic we consider the radion-inflaton potential in the form of 
\beq
\laq{potrad}
V= m_r^2[R] \frac{A^2}{2} +{C_{\rm inf} H_{\rm inf}^2 M_{\rm pl}} A
\eeq
where $A \equiv \sqrt{\frac{3}{2}}M_{\rm pl}\log{(r/R)}$ is the radion with the kinetic normalization.  
$C_{\rm inf}$ denotes the interaction strength between inflaton and $r$ in the 4D Planck unit. 
This should be a good description when $r_{\rm inf}$ is not too much deviated from $R$. 
Then we obtain 
\beq
\frac{1}{r_{\rm inf}}=\frac{1}{R}\exp{\(\frac{C_{\rm inf}H^2_{\rm inf}}{m_r^2[R]}\)}.
\eeq
This gives $r_{\rm inf}\ll R$ if $C_{\rm inf}>0$ and
\beq
H_{\rm inf}\gtrsim {C^{-1/2}_{\rm inf}} m_r[R].
\eeq
 By concerning \Eq{RLB},  the inflation scale turns out to be in the range
\beq
\laq{LHinf}
H_{\rm inf}\gtrsim C_{\rm inf}^{-1/2}10^{10}\GEV \(\frac{10^{15}\GEV}{1/R}\)^2. 
\eeq
This is consistent with \Eq{Hinfup}.

For $r_{\rm inf}$ much larger or smaller than $R$, the expansion in \eq{potrad}
is invalid. One should instead define $A$ with $R$ replaced by $r_{\rm
inf}$ to conduct a similar discussion. The same lower bound on $H_{\rm inf}$
is obtained where $1/R$ is replaced by $1/r_{\rm inf}$.

Notice that we can only have  $1/R\gtrsim 10^{15}\GEV$ to avoid the strong CP problem. 
From Fig.\ref{fig:CPV}, we find that with $1/r_{\rm inf} \sim [1.5-3]\times 10^{16}\GEV$ 
we can have \eq{condmech}. For larger $1/r_{\rm inf}$, the UV instanton contribution is larger than the upper bound of the inflation scale~\eq{Hinfup}. 
Consequently we find that our mechanism works for high scale inflation.

\begin{figure}[!t]
\begin{center}  
      \includegraphics[width=110mm]{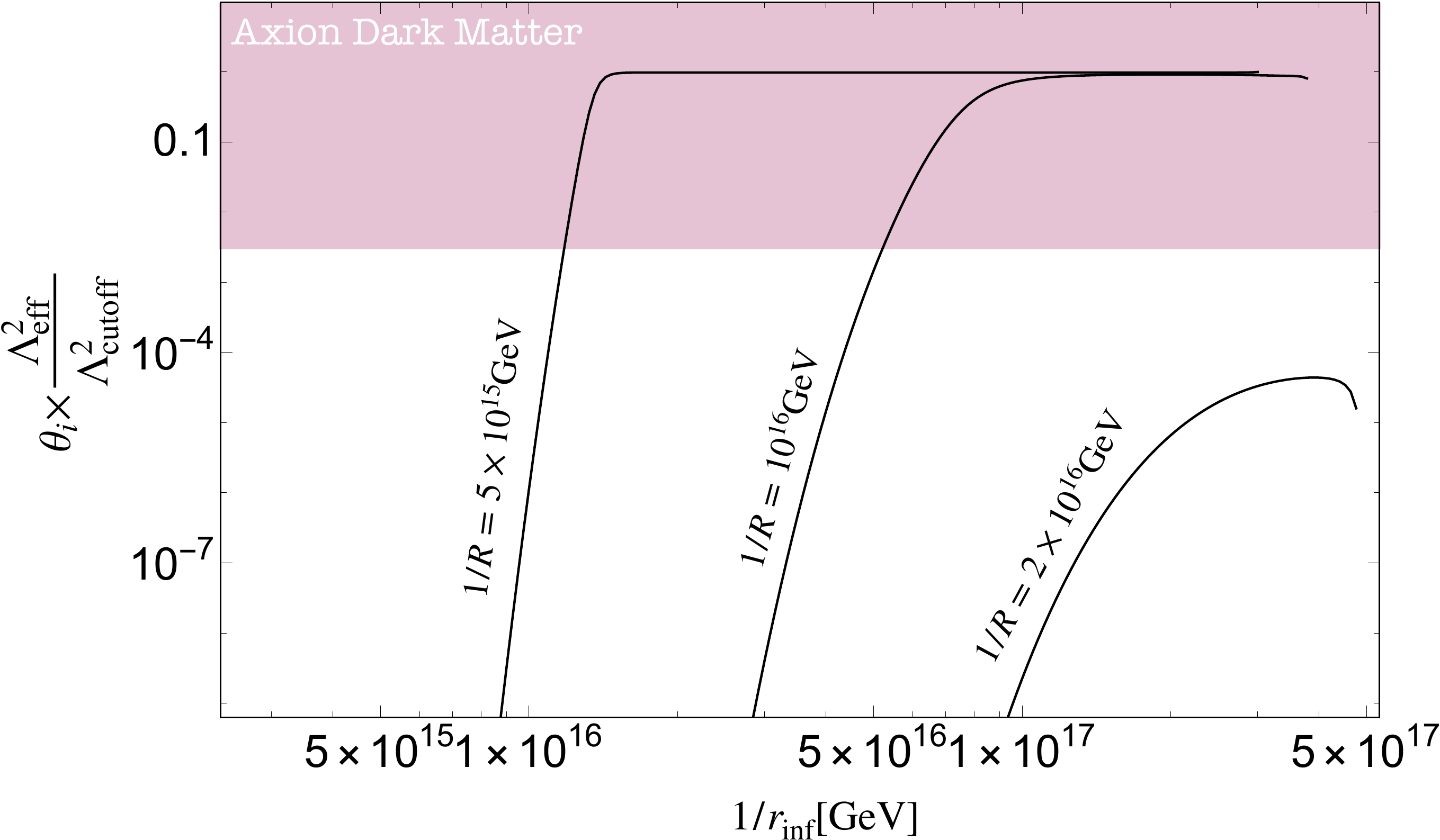}
      \includegraphics[width=105mm]{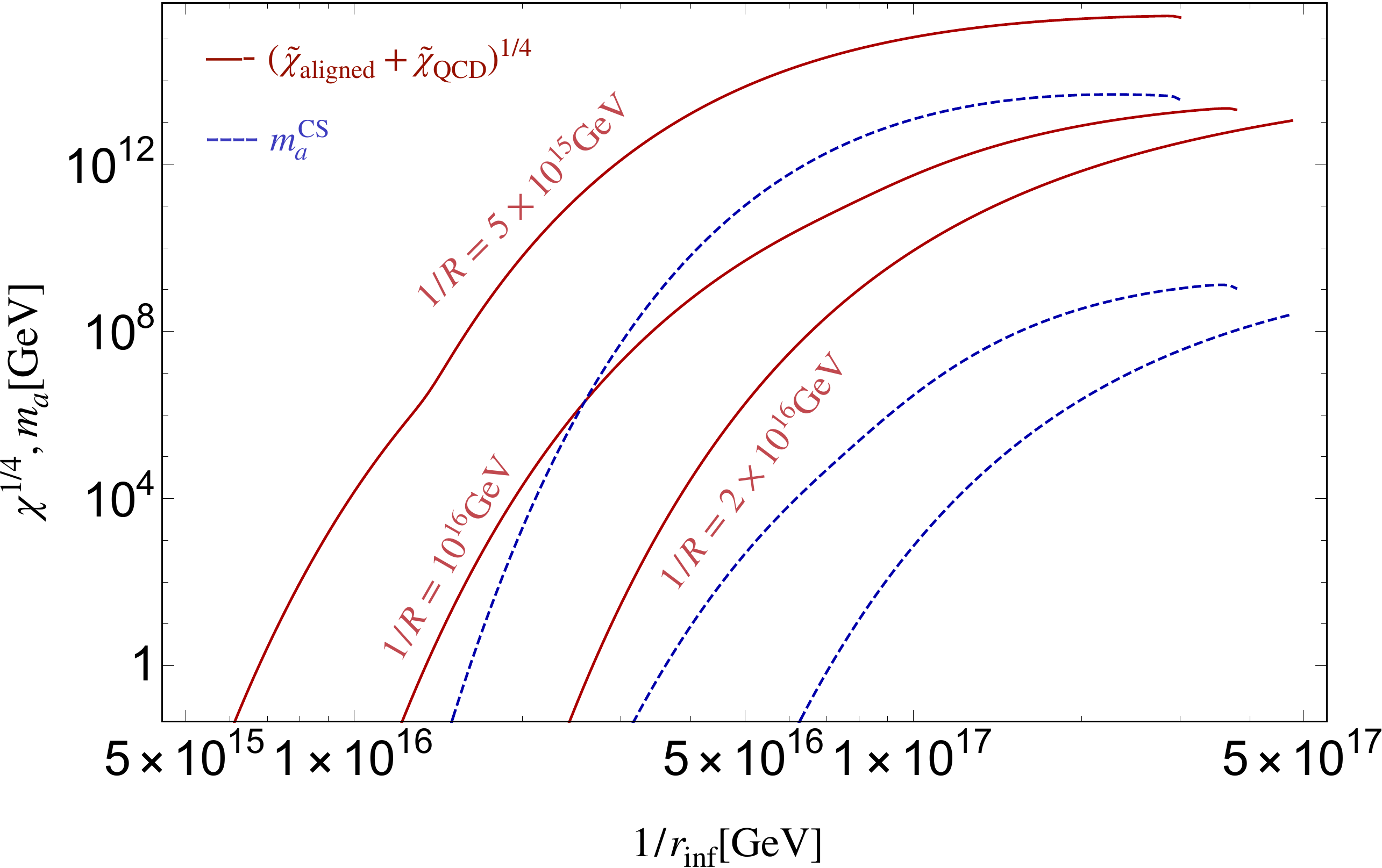}      
      \end{center}
\caption{
The resulting misalignment angle of chiral suppression scenario if axion field is set at the potential minimum at $r=r_{\rm inf}$ during inflation [upper panel]. 
The region can explain the dominant axion dark matter with $\L_{\rm cutoff}/ \L_{\rm eff}\lesssim 1$ is shown in purple. 
The corresponding topological susceptibility$^{1/4}$ [red solid line] and axion mass [blue dashed line] neglecting CPV contribution with general $r$ are also shown [lower panel]. 
}\label{fig:Abundance2} 
\end{figure}

To explain the parameter region, we display  $\theta_i \times \(\frac{ \L_{\rm cutoff}}{\L_{\rm eff}}\)^{-2}$ in the upper panel of Fig.~\ref{fig:Abundance2} for $1/R=5\times 10^{15}\GEV, 10^{16}\GEV \AND 2\times 10^{16}\GEV $ from left to right. 
Here we take $\theta_i=\vev{a/f_a}_r= \tl{\chi}^{\rm CS}_{\rm CPV}/(\tl{\chi}_{\rm aligned}+\tl{\chi}_{\rm QCD})$. Those values are calculated by using \Eqs{aligned} and \eq{MFVCPV}
with multiplying $M_{\rm pl}^4/(r \L_{\rm cutoff}^3)^2$
 and \Eq{QCDr}. 
We also colored the range above $1\%$ which may account for the axion dark matter with $\L_{\rm cutoff}/\L_{\rm eff}\gtrsim \O(10)\%.$  
Larger than a critical point, $\theta_i$ approaches to a constant irrelevant to $r_{\rm inf}$ and $1/R$ when $1/R\lesssim 10^{16}\GEV$. This  $r_{\rm inf}$ insensitive region is what we have been focusing on. Smaller than the critical point we also have a tiny parameter region. 
In the lower panel we present the corresponding $\tl{\chi}_{\rm
aligned} \AND \tl{\chi}_{\rm QCD}$ by red solid and dashed lines
respectively. We can see that $m_a^{\rm CS}[r_{\rm inf}]$ can be larger than $\sim 10^{10}\GEV$, and satisfy both \eq{condmech} and \eq{LHinf} with $1/R\lesssim 10^{16}\GEV$. 
As a result, the parameter region for this scenario is \beq 1/R\sim 10^{15}\GEV\text{-}10^{16}\GEV \AND H_{\rm inf}\sim C_{\rm inf}^{-1/2}10^{10}\GEV\text{-}10^{14}\GEV.\eeq
This region may be searched for in the future EDM experiments and the measurement of the tensor-to-scalar ratio. If the scale is linked to the proton decay operator we may further have implications in the experiments searching for the proton decay (see the next section). 
\\

When $1/R \gg 2\times 10^{16}\GEV$, on the other hand, $\theta_i$ and $m^{\rm CS}_a[r_{\rm inf}]$ are too small for our mechanism to work. However we may have other mechanism to explain the axion dark matter. 
It was shown that if inflation lasts long, $\theta_i$ approaches to the Bunch-Davies distribution~\cite{Graham:2018jyp,Guth:2018hsa}.  In particular, when $ (\tl{\chi}_{\rm QCD}+\tl{\chi}_{\rm aligned})^{1/4}\sim H_{\rm inf}$
we may again obtain dominant axion dark matter. However, since $m^{\rm CS}_a[r_{\rm inf}]\ll H_{\rm inf}$, the axion is almost massless during inflation. 
There is a constraint on the axion quantum fluctuation from
the isocurvature of density perturbation \cite{Akrami:2018odb}, which sets an upper bound on the inflation scale:
\beq
\laq{iso}
H_{\rm inf} \lesssim 10^{9}\GEV \(\frac{f_a}{10^{16}\GEV}\)^{0.408}.
\eeq 
This can be satisfied with $C_{\rm inf}\gtrsim 100$.  
We have checked that when $1/R\lesssim 10^{17}\GEV$ there exists $r_{\rm inf}$ to satisfy $(\tl{\chi}_{\rm QCD}+\tl{\chi}_{\rm aligned})^{1/4}\sim H_{\rm inf}\sim 10^{9}\GEV.$ If $C_{\rm inf}$ is not extremely large, the scenario may be tested in the future CMB observation of the isocurvature perturbation.

Lastly, let us comment  the case without chirality suppressions. In
this case, we can still obtain the axion dark matter similar to the
discussion here. However the inflation scale is extremely low with
$H_{\rm inf}\ll 1\GEV $ (see Fig.\ref{fig:CPV}) and the radion mass
needs to be very light, which may require certain tuning.

\subsection{More unified pictures}
\lac{unif}
More minimally, radion may be the inflaton. In this case it must be away from the potential minimum, $r=R$, during inflation.  
We need to make the radion potential flat enough at around $r\sim 1/\L_{\rm cutoff}$ to drive a hilltop inflation. 
This is always possible if we can write down a general localized potential on a brane. 
In this case, if we can take the potential arbitrary flat, in principle we can have arbitrary small $H_{\rm inf}$. 

An interesting question is whether we can build a GUT model broken at a scale of $1/R$. 
Let us consider a simple GUT gauge group in the bulk and matters are localized on a brane. 
We expect not too different size of small instanton contribution  from that discussed in the main part since the exponential term has the exponent $S_{\rm eff}\sim 2\pi/\a_s\approx 2\pi/\a_{\rm GUT} \approx 8\pi^3 R/g_5^2 $ in the weakly coupled limit, with $\a_{\rm GUT}$ being the 
GUT gauge coupling constant.
Therefore, if the small instanton (from GUT) is still aligned to the IR contribution, we expect the similar scenario. 
An interesting prediction then is the proton decay. The relevant higher dimensional operators, are suppressed by $R^2$. Therefore the decay constant of the axion is linked to the proton decay rate. 
However, the breaking of GUT may induce a small instanton with CPV to the axion potential even in the renormalizable theory (c.f. \cite{Dine:1986bg}). Whether the axion dark matter scenario works would depend on the detail of the GUT models.

\section{Conclusions}

The axion window, $10^{8}~{\rm GeV} \lesssim f_a \lesssim
10^{12}~{\rm GeV}$, has been discussed as the allowed region for the
QCD axion models.
The upper bound, $10^{12}$~GeV, is put by cosmological considerations
where the misalignment of the axion value in the early Universe
produces the axion energy density at later time as oscillations about
the true minimum of the potential.
This upper bound has given a theoretical challenge to lower the decay
constant much below the scale of grand unification or quantum gravity.

The challenge should however be considered carefully. In the UV
physics, there can be unknown significant contribution to the axion
mass which may be aligned to the low energy QCD contributions so that
the strong CP problem is still solved.
Also, the UV physics may modify the axion potential in the early
Universe so that the upper bound itself is not reliable.

We discussed that the UV contribution, if it is dominated by high
scale physics such as the Planck scale, is severely constrained by a
new CP violation caused by the combination of instantons and higher
dimensional operators.
Especially, in the model where the axion arises from a gauge field in
the extra dimension, such as string axions, the new CP problem
excludes the possibility for the UV contributions to overwhelm the
ordinary low energy QCD contributions.
The consistent scenario is only possible for the size of the extra
dimension, i.e., the axion decay constant, to be larger than about
$10^{15-16}$~GeV which is beyond the axion window.

It is however important to realize that the UV contribution can be
much larger during the cosmological inflation as the size of the extra
dimension can be different from the current one.
By considering the enhancement of the QCD coupling during the
inflation, the minimum of the axion potential is located near the
current minimum in a wide range of parameters, so that a small
misalignment angle is realized. Even for a large decay constant, such
as $f_a \sim 10^{15-16}~{\rm GeV}$, axion abundance can naturally be that
of dark matter of the Universe while the strong CP problem is still
solved. This discussion opens up the possibility of natural string or
GUT axion scenarios where inflation dynamics is tied to the size of
the compactified directions.
The upper bound on the axion window should not be taken so seriously
in string models.

\section*{Acknowledgements} 
W.Y. thanks the KEK theory group for the kind hospitality when this work is done.  
R.K. thanks Norimi Yokozaki for useful discussions. 
This work is supported by JSPS KAKENHI Grant-in-Aid for Scientific
Research (Nos.~19H00689 [RK] and 19H05810 [WY]), MEXT KAKENHI Grant-in-Aid for Scientific
Research on Innovative Areas (No.~18H05542 [RK]).

\appendix
\section{A model of heavy axion from accidental PQ symmetry}
\lac{model}
To have a closer look of the quality problem raised here. Let us build a new kind of heavy axion model, and discuss the CP violation. 
Let us consider again $Z_N$ (gauge) symmetry to solve the ordinary quality problem and ``define" the PQ symmetry. We consider $N$ fundamental quark pairs, $\bar{Q}_{\rm PQ}^a, {Q}_{\rm PQ}^a$ with interaction of 
\beq \laq{PQ2}{\cal L}\supset - \F_{\rm PQ}\sum_{a}y_\F^{(a)}\bar{Q}^a_{\rm PQ}Q^a_{\rm PQ}.\eeq
 Under $Z_N$, $\F_{\rm PQ}^*, \AND \bar{Q}_{\rm PQ}^a$  transform with a phase of $-2\pi/N$, $Q_{\rm PQ}$ is a singlet. The quark masses are assumed to be around the PQ scale $f_a$, i.e. $y^{(a)}_\F\sim 1.$ 
In addition, we introduce $N_\f$ of fundamental colored scalars, $\f_i$, around the mass scale, $m_\f$. 
Then \Eq{vac} can be denoted as the integral by parts as 
\begin{align}
{\cal L}^{\rm (vac)}_{\rm eff}&\sim 
  \int^{1/ m_{\f}}_{1/f_a}{\frac{d \rho' }{\rho'^5} e^{-S^{(1)}_{\rm eff}[1/\rho']}  \
  F^{\rm (1)}[\rho']
   \det{\frac{Y_{u}}{4\pi}} \det{\frac{Y_{d}}{4\pi}} } \cos{\(\frac{N a}{f_a}\)}\\
   &+
  \int^{1/f_a}_{1/\L_{\rm cutoff}}{\frac{d \rho' }{\rho'^5} \rho'^{N} f_a^{N} e^{-S^{(2)}_{\rm eff}[1/\rho']}  \
  F^{\rm (2)}[\rho']
   \det{\frac{Y_{u}}{4\pi}} \det{\frac{Y_{d}}{4\pi}} } \cos{\(\frac{N a}{f_a}\)}
   .\laq{int2}
\end{align}
If the number of $\f$, $N_\f$ is large enough, the gauge coupling is no longer asymptotically free. 
Then we can obtain
\beq
\exp{[-S^{(1,2)}_{\rm eff}]}\propto \rho'^{b_{1,2}}
\eeq
with $b_1=b_0 - N_{\f}/6$ 
and $b_2=b_0 - N_{\f}/6-2N/3$.
Neglecting the $\rho'$ dependence in $F^{(i)}[\rho'],$ 
the integral may dominates around 
\beq
\rho_{\rm case 1} \sim1/m_\f, \rho_{\rm case 2}\sim 1/f_a, \rho_{\rm case 3}\sim 1/\L_{\rm cutoff},\eeq
depending on $N_\f\AND N.$
The first case is not at all problematic because $S_{\rm eff}[1/\rho]\approx 2\pi /\a_{\rm SM}[1/\rho],$ which is around the suppressed SM contribution. 
The third case may induce an additional potential to axion but it can be suppressed by assuming a large $N$ as the ordinary solution to the quality problem,
i.e. in this case the term is suppressed by $f_a^N/\L_{\rm cutoff}^{N-4} $. 
This also says that with increasing $N$ the case 3 approaches to the case 2 since the integral in \eq{int2} tends to dominate at IR. 

The case 2, the integral dominates at  $\rho\sim 1/f_a,$ may generate additional axion mass. 
Then we get the small instanton contribution as
\beq
\e \chi_0\sim   10^{-23} f_a^4\times \(\frac{2\pi}{\a_s[f_a]}\)^6 e^{-\frac{2\pi}{\alpha_s[f_a]}}. 
\eeq
with 
\beq 
e^{-\frac{2\pi}{\alpha_s[f_a]}}\sim\(\frac{0.05\GEV}{m_\f}\)^{b_0} \(\frac{m_\f}{f_a}\)^{b_1}. 
\eeq
If $-b_1+4 \geq 0$, this contribution is enhanced. For instance with $N_\f=60$ \AND $f_a=8\times 10^8$, we get $\e\sim 10^5$ and obtain the heavy axion. 
Here $\a_s[f_a]\sim 0.12.$ We do not need to care the UV contribution since we can take $N$ large enough to suppress it. 
But we should make sure that before the coupling blows up, there is some UV completion appear (by integrating out which we may have another small instanton effect but again it is suppressed as long as $N$ is large). 
Then  \eq{pot} suggests $1/\rho \sim f_a\lesssim 10^{10}\GEV$ when $\e \gg 1.$ 
The parameter region of the axion dark matter of this model is shown by the red line in Fig.~\ref{fig:Mpl2}.


\begin{thebibliography}{99}
\bibitem{Peccei:1977hh}
R.~D.~Peccei and H.~R.~Quinn,
Phys. Rev. Lett. \textbf{38}, 1440-1443 (1977)
doi:10.1103/PhysRevLett.38.1440

\bibitem{Peccei:1977ur}
R.~D.~Peccei and H.~R.~Quinn,
Phys. Rev. D \textbf{16}, 1791-1797 (1977)
doi:10.1103/PhysRevD.16.1791

\bibitem{Weinberg:1977ma}
S.~Weinberg,
Phys. Rev. Lett. \textbf{40}, 223-226 (1978)
doi:10.1103/PhysRevLett.40.223

\bibitem{Wilczek:1977pj}
F.~Wilczek,
Phys. Rev. Lett. \textbf{40}, 279-282 (1978)
doi:10.1103/PhysRevLett.40.279

\bibitem{Abbott:1982af}
L.~F.~Abbott and P.~Sikivie,
Phys. Lett. B \textbf{120}, 133-136 (1983)
doi:10.1016/0370-2693(83)90638-X

\bibitem{Preskill:1982cy}
J.~Preskill, M.~B.~Wise and F.~Wilczek,
Phys. Lett. B \textbf{120}, 127-132 (1983)
doi:10.1016/0370-2693(83)90637-8

\bibitem{Dine:1982ah}
M.~Dine and W.~Fischler,
Phys. Lett. B \textbf{120}, 137-141 (1983)
doi:10.1016/0370-2693(83)90639-1

\bibitem{Witten:1984dg}
E.~Witten,
Phys. Lett. B \textbf{149}, 351-356 (1984)
doi:10.1016/0370-2693(84)90422-2

\bibitem{Svrcek:2006yi}
P.~Svrcek and E.~Witten,
JHEP \textbf{06}, 051 (2006)
doi:10.1088/1126-6708/2006/06/051
[arXiv:hep-th/0605206 [hep-th]].

\bibitem{Conlon:2006tq}
J.~P.~Conlon,
JHEP \textbf{05}, 078 (2006)
doi:10.1088/1126-6708/2006/05/078
[arXiv:hep-th/0602233 [hep-th]].

\bibitem{Arvanitaki:2009fg}
A.~Arvanitaki, S.~Dimopoulos, S.~Dubovsky, N.~Kaloper and J.~March-Russell,
Phys. Rev. D \textbf{81}, 123530 (2010)
doi:10.1103/PhysRevD.81.123530
[arXiv:0905.4720 [hep-th]].

\bibitem{Acharya:2010zx}
B.~S.~Acharya, K.~Bobkov and P.~Kumar,
JHEP \textbf{11}, 105 (2010)
doi:10.1007/JHEP11(2010)105
[arXiv:1004.5138 [hep-th]].

\bibitem{Higaki:2011me}
T.~Higaki and T.~Kobayashi,
Phys. Rev. D \textbf{84}, 045021 (2011)
doi:10.1103/PhysRevD.84.045021
[arXiv:1106.1293 [hep-th]].

\bibitem{Cicoli:2012sz}
M.~Cicoli, M.~Goodsell and A.~Ringwald,
JHEP \textbf{10}, 146 (2012)
doi:10.1007/JHEP10(2012)146
[arXiv:1206.0819 [hep-th]].

\bibitem{Demirtas:2018akl}
M.~Demirtas, C.~Long, L.~McAllister and M.~Stillman,
JHEP \textbf{04}, 138 (2020)
doi:10.1007/JHEP04(2020)138
[arXiv:1808.01282 [hep-th]].

\bibitem{Jaeckel:2010ni}
J.~Jaeckel and A.~Ringwald,
Ann. Rev. Nucl. Part. Sci. \textbf{60}, 405-437 (2010)
doi:10.1146/annurev.nucl.012809.104433
[arXiv:1002.0329 [hep-ph]].

\bibitem{Ringwald:2012hr}
A.~Ringwald,
Phys. Dark Univ. \textbf{1}, 116-135 (2012)
doi:10.1016/j.dark.2012.10.008
[arXiv:1210.5081 [hep-ph]].

\bibitem{Arias:2012az}
P.~Arias, D.~Cadamuro, M.~Goodsell, J.~Jaeckel, J.~Redondo and A.~Ringwald,
JCAP \textbf{06}, 013 (2012)
doi:10.1088/1475-7516/2012/06/013
[arXiv:1201.5902 [hep-ph]].

\bibitem{Graham:2015ouw}
P.~W.~Graham, I.~G.~Irastorza, S.~K.~Lamoreaux, A.~Lindner and K.~A.~van Bibber,
Ann. Rev. Nucl. Part. Sci. \textbf{65}, 485-514 (2015)
doi:10.1146/annurev-nucl-102014-022120
[arXiv:1602.00039 [hep-ex]].

\bibitem{Marsh:2015xka}
D.~J.~E.~Marsh,
Phys. Rept. \textbf{643}, 1-79 (2016)
doi:10.1016/j.physrep.2016.06.005
[arXiv:1510.07633 [astro-ph.CO]].

\bibitem{Irastorza:2018dyq}
I.~G.~Irastorza and J.~Redondo,
Prog. Part. Nucl. Phys. \textbf{102}, 89-159 (2018)
doi:10.1016/j.ppnp.2018.05.003
[arXiv:1801.08127 [hep-ph]].

\bibitem{DiLuzio:2020wdo}
L.~Di Luzio, M.~Giannotti, E.~Nardi and L.~Visinelli,
Phys. Rept. \textbf{870}, 1-117 (2020)
doi:10.1016/j.physrep.2020.06.002
[arXiv:2003.01100 [hep-ph]].

\bibitem{Misner:1957mt}
C.~W.~Misner and J.~A.~Wheeler,
Annals Phys. \textbf{2}, 525-603 (1957)
doi:10.1016/0003-4916(57)90049-0

\bibitem{Barr:1992qq}
S.~M.~Barr and D.~Seckel,
Phys. Rev. D \textbf{46}, 539-549 (1992)
doi:10.1103/PhysRevD.46.539

\bibitem{Banks:2010zn}
T.~Banks and N.~Seiberg,
Phys. Rev. D \textbf{83}, 084019 (2011)
doi:10.1103/PhysRevD.83.084019
[arXiv:1011.5120 [hep-th]].

\bibitem{Alvey:2020nyh}
J.~Alvey and M.~Escudero,
JHEP \textbf{01}, 032 (2021)
doi:10.1007/JHEP01(2021)032
[arXiv:2009.03917 [hep-ph]].

\bibitem{Chun:1992bn}
E.~J.~Chun and A.~Lukas,
Phys. Lett. B \textbf{297}, 298-304 (1992)
doi:10.1016/0370-2693(92)91266-C
[arXiv:hep-ph/9209208 [hep-ph]].

\bibitem{BasteroGil:1997vn}
M.~Bastero-Gil and S.~F.~King,
Phys. Lett. B \textbf{423}, 27-34 (1998)
doi:10.1016/S0370-2693(98)00124-5
[arXiv:hep-ph/9709502 [hep-ph]].

\bibitem{Babu:2002ic}
K.~S.~Babu, I.~Gogoladze and K.~Wang,
Phys. Lett. B \textbf{560}, 214-222 (2003)
doi:10.1016/S0370-2693(03)00411-8
[arXiv:hep-ph/0212339 [hep-ph]].

\bibitem{Fukuda:2017ylt}
H.~Fukuda, M.~Ibe, M.~Suzuki and T.~T.~Yanagida,
Phys. Lett. B \textbf{771}, 327-331 (2017)
doi:10.1016/j.physletb.2017.05.071
[arXiv:1703.01112 [hep-ph]].

\bibitem{Duerr:2017amf}
M.~Duerr, K.~Schmidt-Hoberg and J.~Unwin,
Phys. Lett. B \textbf{780}, 553-556 (2018)
doi:10.1016/j.physletb.2018.03.054
[arXiv:1712.01841 [hep-ph]].

\bibitem{Bonnefoy:2018ibr}
Q.~Bonnefoy, E.~Dudas and S.~Pokorski,
Eur. Phys. J. C \textbf{79}, no.1, 31 (2019)
doi:10.1140/epjc/s10052-018-6528-z
[arXiv:1804.01112 [hep-ph]].

\bibitem{Redi:2016esr}
M.~Redi and R.~Sato,
JHEP \textbf{05}, 104 (2016)
doi:10.1007/JHEP05(2016)104
[arXiv:1602.05427 [hep-ph]].

\bibitem{Darme:2021cxx}
L.~Darm\'e and E.~Nardi,
[arXiv:2102.05055 [hep-ph]].

\bibitem{Nakai:2021nyf}
Y.~Nakai and M.~Suzuki,
[arXiv:2102.01329 [hep-ph]].

\bibitem{Cheng:2001ys}
H.~C.~Cheng and D.~E.~Kaplan,
[arXiv:hep-ph/0103346 [hep-ph]].

\bibitem{Choi:2003wr}
K.~w.~Choi,
Phys. Rev. Lett. \textbf{92}, 101602 (2004)
doi:10.1103/PhysRevLett.92.101602
[arXiv:hep-ph/0308024 [hep-ph]].

\bibitem{Marsh:2019bjr}
D.~J.~E.~Marsh and W.~Yin,
JHEP \textbf{01}, 169 (2021)
doi:10.1007/JHEP01(2021)169
[arXiv:1912.08188 [hep-ph]].

\bibitem{Agrawal:2017ksf}
P.~Agrawal and K.~Howe,
JHEP \textbf{12}, 029 (2018)
doi:10.1007/JHEP12(2018)029
[arXiv:1710.04213 [hep-ph]].

\bibitem{Csaki:2019vte}
C.~Cs\'aki, M.~Ruhdorfer and Y.~Shirman,
JHEP \textbf{04}, 031 (2020)
doi:10.1007/JHEP04(2020)031
[arXiv:1912.02197 [hep-ph]].

\bibitem{Gherghetta:2020ofz}
T.~Gherghetta and M.~D.~Nguyen,
JHEP \textbf{12}, 094 (2020)
doi:10.1007/JHEP12(2020)094
[arXiv:2007.10875 [hep-ph]].

\bibitem{Gupta:2020vxb}
R.~S.~Gupta, V.~V.~Khoze and M.~Spannowsky,
[arXiv:2012.00017 [hep-ph]].

\bibitem{Poppitz:2002ac}
E.~Poppitz and Y.~Shirman,
JHEP \textbf{07}, 041 (2002)
doi:10.1088/1126-6708/2002/07/041
[arXiv:hep-th/0204075 [hep-th]].

\bibitem{Gherghetta:2020keg}
T.~Gherghetta, V.~V.~Khoze, A.~Pomarol and Y.~Shirman,
JHEP \textbf{03}, 063 (2020)
doi:10.1007/JHEP03(2020)063
[arXiv:2001.05610 [hep-ph]].

\bibitem{Callan:1977gz}
C.~G.~Callan, Jr., R.~F.~Dashen and D.~J.~Gross,
Phys. Rev. D \textbf{17}, 2717 (1978)
doi:10.1103/PhysRevD.17.2717

\bibitem{Holdom:1982ex}
B.~Holdom and M.~E.~Peskin,
Nucl. Phys. B \textbf{208}, 397-412 (1982)
doi:10.1016/0550-3213(82)90228-0

\bibitem{Dine:1986bg}
M.~Dine and N.~Seiberg,
Nucl. Phys. B \textbf{273}, 109-124 (1986)
doi:10.1016/0550-3213(86)90043-X

\bibitem{Flynn:1987rs}
J.~M.~Flynn and L.~Randall,
Nucl. Phys. B \textbf{293}, 731-739 (1987)
doi:10.1016/0550-3213(87)90089-7

\bibitem{tHooft:1976snw}
G.~'t Hooft,
Phys. Rev. D \textbf{14}, 3432-3450 (1976)
[erratum: Phys. Rev. D \textbf{18}, 2199 (1978)]
doi:10.1103/PhysRevD.14.3432

\bibitem{Kim:1979if}
J.~E.~Kim,
Phys. Rev. Lett. \textbf{43}, 103 (1979)
doi:10.1103/PhysRevLett.43.103

\bibitem{Shifman:1979if}
M.~A.~Shifman, A.~I.~Vainshtein and V.~I.~Zakharov,
Nucl. Phys. B \textbf{166}, 493-506 (1980)
doi:10.1016/0550-3213(80)90209-6

\bibitem{Dine:1981rt}
M.~Dine, W.~Fischler and M.~Srednicki,
Phys. Lett. B \textbf{104}, 199-202 (1981)
doi:10.1016/0370-2693(81)90590-6

\bibitem{Zhitnitsky:1980tq}
A.~R.~Zhitnitsky,
Sov. J. Nucl. Phys. \textbf{31}, 260 (1980)

\bibitem{Baker:2006ts}
C.~A.~Baker, D.~D.~Doyle, P.~Geltenbort, K.~Green, M.~G.~D.~van der Grinten, P.~G.~Harris, P.~Iaydjiev, S.~N.~Ivanov, D.~J.~R.~May and J.~M.~Pendlebury, \textit{et al.}
Phys. Rev. Lett. \textbf{97}, 131801 (2006)
doi:10.1103/PhysRevLett.97.131801
[arXiv:hep-ex/0602020 [hep-ex]].

\bibitem{Afach:2015sja}
J.~M.~Pendlebury, S.~Afach, N.~J.~Ayres, C.~A.~Baker, G.~Ban, G.~Bison, K.~Bodek, M.~Burghoff, P.~Geltenbort and K.~Green, \textit{et al.}
Phys. Rev. D \textbf{92}, no.9, 092003 (2015)
doi:10.1103/PhysRevD.92.092003
[arXiv:1509.04411 [hep-ex]].

\bibitem{Pospelov:2005pr}
M.~Pospelov and A.~Ritz,
Annals Phys. \textbf{318}, 119-169 (2005)
doi:10.1016/j.aop.2005.04.002
[arXiv:hep-ph/0504231 [hep-ph]].

\bibitem{Dragos:2019oxn}
J.~Dragos, T.~Luu, A.~Shindler, J.~de Vries and A.~Yousif,
Phys. Rev. C \textbf{103}, no.1, 015202 (2021)
doi:10.1103/PhysRevC.103.015202
[arXiv:1902.03254 [hep-lat]].

\bibitem{Anastassopoulos:2015ura}
V.~Anastassopoulos, S.~Andrianov, R.~Baartman, M.~Bai, S.~Baessler, J.~Benante, M.~Berz, M.~Blaskiewicz, T.~Bowcock and K.~Brown, \textit{et al.}
Rev. Sci. Instrum. \textbf{87}, no.11, 115116 (2016)
doi:10.1063/1.4967465
[arXiv:1502.04317 [physics.acc-ph]].

\bibitem{Randall:1992ut}
L.~Randall,
Phys. Lett. B \textbf{284}, 77-80 (1992)
doi:10.1016/0370-2693(92)91928-3

\bibitem{DiLuzio:2017tjx}
L.~Di Luzio, E.~Nardi and L.~Ubaldi,
Phys. Rev. Lett. \textbf{119}, no.1, 011801 (2017)
doi:10.1103/PhysRevLett.119.011801
[arXiv:1704.01122 [hep-ph]].

\bibitem{Lillard:2018fdt}
B.~Lillard and T.~M.~P.~Tait,
JHEP \textbf{11}, 199 (2018)
doi:10.1007/JHEP11(2018)199
[arXiv:1811.03089 [hep-ph]].

\bibitem{Lee:2018yak}
H.~S.~Lee and W.~Yin,
Phys. Rev. D \textbf{99}, no.1, 015041 (2019)
doi:10.1103/PhysRevD.99.015041
[arXiv:1811.04039 [hep-ph]].

\bibitem{Ardu:2020qmo}
M.~Ardu, L.~Di Luzio, G.~Landini, A.~Strumia, D.~Teresi and J.~W.~Wang,
JHEP \textbf{11}, 090 (2020)
doi:10.1007/JHEP11(2020)090
[arXiv:2007.12663 [hep-ph]].

\bibitem{Yin:2020dfn}
W.~Yin,
JHEP \textbf{10}, 032 (2020)
doi:10.1007/JHEP10(2020)032
[arXiv:2007.13320 [hep-ph]].

\bibitem{Yamada:2021uze}
M.~Yamada and T.~T.~Yanagida,
[arXiv:2101.10350 [hep-ph]].

\bibitem{Agrawal:2017evu}
P.~Agrawal and K.~Howe,
JHEP \textbf{12}, 035 (2018)
doi:10.1007/JHEP12(2018)035
[arXiv:1712.05803 [hep-ph]].

\bibitem{Mayle:1987as}
R.~Mayle, J.~R.~Wilson, J.~R.~Ellis, K.~A.~Olive, D.~N.~Schramm and G.~Steigman,
Phys. Lett. B \textbf{203}, 188-196 (1988)
doi:10.1016/0370-2693(88)91595-X

\bibitem{Raffelt:1987yt}
G.~Raffelt and D.~Seckel,
Phys. Rev. Lett. \textbf{60}, 1793 (1988)
doi:10.1103/PhysRevLett.60.1793

\bibitem{Turner:1987by}
M.~S.~Turner,
Phys. Rev. Lett. \textbf{60}, 1797 (1988)
doi:10.1103/PhysRevLett.60.1797

\bibitem{Chang:2018rso}
J.~H.~Chang, R.~Essig and S.~D.~McDermott,
JHEP \textbf{09}, 051 (2018)
doi:10.1007/JHEP09(2018)051
[arXiv:1803.00993 [hep-ph]].

\bibitem{Berkowitz:2015aua}
E.~Berkowitz, M.~I.~Buchoff and E.~Rinaldi,
Phys. Rev. D \textbf{92}, no.3, 034507 (2015)
doi:10.1103/PhysRevD.92.034507
[arXiv:1505.07455 [hep-ph]].

\bibitem{Kitano:2015fla}
R.~Kitano and N.~Yamada,
JHEP \textbf{10}, 136 (2015)
doi:10.1007/JHEP10(2015)136
[arXiv:1506.00370 [hep-ph]].

\bibitem{Borsanyi:2015cka}
S.~Borsanyi, M.~Dierigl, Z.~Fodor, S.~D.~Katz, S.~W.~Mages, D.~Nogradi, J.~Redondo, A.~Ringwald and K.~K.~Szabo,
Phys. Lett. B \textbf{752}, 175-181 (2016)
doi:10.1016/j.physletb.2015.11.020
[arXiv:1508.06917 [hep-lat]].

\bibitem{Frison:2016vuc}
J.~Frison, R.~Kitano, H.~Matsufuru, S.~Mori and N.~Yamada,
JHEP \textbf{09}, 021 (2016)
doi:10.1007/JHEP09(2016)021
[arXiv:1606.07175 [hep-lat]].

\bibitem{Borsanyi:2016ksw}
S.~Borsanyi, Z.~Fodor, J.~Guenther, K.~H.~Kampert, S.~D.~Katz, T.~Kawanai, T.~G.~Kovacs, S.~W.~Mages, A.~Pasztor and F.~Pittler, \textit{et al.}
Nature \textbf{539}, no.7627, 69-71 (2016)
doi:10.1038/nature20115
[arXiv:1606.07494 [hep-lat]].

\bibitem{Ho:2019ayl}
S.~Y.~Ho, F.~Takahashi and W.~Yin,
JHEP \textbf{04}, 149 (2019)
doi:10.1007/JHEP04(2019)149
[arXiv:1901.01240 [hep-ph]].

\bibitem{Ballesteros:2016xej}
G.~Ballesteros, J.~Redondo, A.~Ringwald and C.~Tamarit,
JCAP \textbf{08}, 001 (2017)
doi:10.1088/1475-7516/2017/08/001
[arXiv:1610.01639 [hep-ph]].

\bibitem{Aghanim:2018eyx}
N.~Aghanim \textit{et al.} [Planck],
Astron. Astrophys. \textbf{641}, A6 (2020)
doi:10.1051/0004-6361/201833910
[arXiv:1807.06209 [astro-ph.CO]].

\bibitem{Essig:2013goa}
R.~Essig, E.~Kuflik, S.~D.~McDermott, T.~Volansky and K.~M.~Zurek,
JHEP \textbf{11}, 193 (2013)
doi:10.1007/JHEP11(2013)193
[arXiv:1309.4091 [hep-ph]].

\bibitem{Moroi:2020has}
T.~Moroi and W.~Yin,
[arXiv:2011.09475 [hep-ph]].

\bibitem{Daido:2017wwb}
R.~Daido, F.~Takahashi and W.~Yin,
JCAP \textbf{05}, 044 (2017)
doi:10.1088/1475-7516/2017/05/044
[arXiv:1702.03284 [hep-ph]].

\bibitem{Co:2017mop}
R.~T.~Co, L.~J.~Hall and K.~Harigaya,
Phys. Rev. Lett. \textbf{120}, no.21, 211602 (2018)
doi:10.1103/PhysRevLett.120.211602
[arXiv:1711.10486 [hep-ph]].

\bibitem{Co:2018mho}
R.~T.~Co, E.~Gonzalez and K.~Harigaya,
JHEP \textbf{05}, 163 (2019)
doi:10.1007/JHEP05(2019)163
[arXiv:1812.11192 [hep-ph]].

\bibitem{Takahashi:2019pqf}
F.~Takahashi and W.~Yin,
JHEP \textbf{10}, 120 (2019)
doi:10.1007/JHEP10(2019)120
[arXiv:1908.06071 [hep-ph]].

\bibitem{Takahashi:2019qmh}
F.~Takahashi and W.~Yin,
JHEP \textbf{07}, 095 (2019)
doi:10.1007/JHEP07(2019)095
[arXiv:1903.00462 [hep-ph]].

\bibitem{Kobayashi:2019eyg}
T.~Kobayashi and L.~Ubaldi,
JHEP \textbf{08}, 147 (2019)
doi:10.1007/JHEP08(2019)147
[arXiv:1907.00984 [hep-ph]].

\bibitem{Nakagawa:2020eeg}
S.~Nakagawa, F.~Takahashi and W.~Yin,
JCAP \textbf{05}, 004 (2020)
doi:10.1088/1475-7516/2020/05/004
[arXiv:2002.12195 [hep-ph]].

\bibitem{Huang:2020etx}
J.~Huang, A.~Madden, D.~Racco and M.~Reig,
JHEP \textbf{10}, 143 (2020)
doi:10.1007/JHEP10(2020)143
[arXiv:2006.07379 [hep-ph]].

\bibitem{Nakagawa:2020zjr}
S.~Nakagawa, F.~Takahashi and M.~Yamada,
[arXiv:2012.13592 [hep-ph]].

\bibitem{Vafa:1984xg}
C.~Vafa and E.~Witten,
Phys. Rev. Lett. \textbf{53}, 535 (1984)
doi:10.1103/PhysRevLett.53.535

\bibitem{Choi:1996fs}
K.~Choi, H.~B.~Kim and J.~E.~Kim,
Nucl. Phys. B \textbf{490}, 349-364 (1997)
doi:10.1016/S0550-3213(97)00066-7
[arXiv:hep-ph/9606372 [hep-ph]].

\bibitem{Dvali:1995ce}
G.~R.~Dvali,
[arXiv:hep-ph/9505253 [hep-ph]].

\bibitem{Banks:1996ea}
T.~Banks and M.~Dine,
Nucl. Phys. B \textbf{505}, 445-460 (1997)
doi:10.1016/S0550-3213(97)00413-6
[arXiv:hep-th/9608197 [hep-th]].

\bibitem{Jeong:2013xta}
K.~S.~Jeong and F.~Takahashi,
Phys. Lett. B \textbf{727}, 448-451 (2013)
doi:10.1016/j.physletb.2013.10.061
[arXiv:1304.8131 [hep-ph]].

\bibitem{Matsui:2020wfx}
H.~Matsui, F.~Takahashi and W.~Yin,
JHEP \textbf{05}, 154 (2020)
doi:10.1007/JHEP05(2020)154
[arXiv:2001.04464 [hep-ph]].

\bibitem{Buen-Abad:2019uoc}
M.~A.~Buen-Abad and J.~Fan,
JHEP \textbf{12}, 161 (2019)
doi:10.1007/JHEP12(2019)161
[arXiv:1911.05737 [hep-ph]].

\bibitem{Linde:1996cx}
A.~D.~Linde,
Phys. Rev. D \textbf{53}, 4129-4132 (1996)
doi:10.1103/PhysRevD.53.R4129
[arXiv:hep-th/9601083 [hep-th]].

\bibitem{Nakayama:2011wqa}
K.~Nakayama, F.~Takahashi and T.~T.~Yanagida,
Phys. Rev. D \textbf{84}, 123523 (2011)
doi:10.1103/PhysRevD.84.123523
[arXiv:1109.2073 [hep-ph]].

\bibitem{Akrami:2018odb}
Y.~Akrami \textit{et al.} [Planck],
Astron. Astrophys. \textbf{641}, A10 (2020)
doi:10.1051/0004-6361/201833887
[arXiv:1807.06211 [astro-ph.CO]].

\bibitem{Ponton:2001hq}
E.~Ponton and E.~Poppitz,
JHEP \textbf{06}, 019 (2001)
doi:10.1088/1126-6708/2001/06/019
[arXiv:hep-ph/0105021 [hep-ph]].

\bibitem{Graham:2018jyp}
P.~W.~Graham and A.~Scherlis,
Phys. Rev. D \textbf{98}, no.3, 035017 (2018)
doi:10.1103/PhysRevD.98.035017
[arXiv:1805.07362 [hep-ph]].

\bibitem{Guth:2018hsa}
F.~Takahashi, W.~Yin and A.~H.~Guth,
Phys. Rev. D \textbf{98}, no.1, 015042 (2018)
doi:10.1103/PhysRevD.98.015042
[arXiv:1805.08763 [hep-ph]].
\end{thebibliography}
\end{document}